\documentstyle[aps,epsf,epsfig,floats]{revtex} 
\begin{document}
\preprint{\vbox{\hbox{DOE/ER/40762-010}\hbox{UM PP\#02-012}}}
\title{The Nucleon-Nucleon Interaction and Large $N_c$ QCD}
\author{Manoj K.~Banerjee, Thomas D.~Cohen and Boris A.~Gelman}
\address{Department of Physics, University of Maryland, College Park, MD 
20742-4111.} 
\maketitle
\begin{abstract}
The nature of the nonrelativistic nucleon-nucleon potential in the large 
$N_c$ limit is discussed. In particular, we address the consistency of the 
meson exchange picture of nucleon interactions. It is shown that the
nonrelativistic nucleon-nucleon potential extracted from the Feynmann graphs
up to and including two-meson exchange diagrams satisfies the spin-flavor
counting rules of Kaplan and Savage, and Kaplan and Manohar, provided the
nucleon momenta is of order $N_{c}^{0}$. The key to this is a cancelation of
the retardation effect of the box graphs against the contributions of the
crossed-box diagram. The consistency requires including $\Delta$ as an
intermediate state.
\end{abstract}
%\narrowtext

\section{Introduction}

One of the most fundamental problems in nuclear physics is to
understand how low-energy nucleon-nucleon interactions arise from
the underlying quark-gluon interactions. Unfortunately, for some
time to come QCD is likely to remain computationally intractable in
this regime in the sense that one will not be able to directly
predict the values of low energy nuclear physics observables by
calculations based solely on the QCD Lagrangian. Nevertheless one
might hope to be able to deduce some qualitative (or perhaps
semi-quantitative) features of nucleon-nucleon interactions from our
knowledge of QCD. The known simplifications of certain aspects of
QCD in the large $N_c$ limit could provide such a tool 
\cite{LN1,LN2,SF1,SF2,SF3}. Indeed,
several years ago it was proposed that the spin-flavor structure of the
dominant terms in the nucleon-nucleon potential can be understood in
terms of large $N_c$ QCD \cite{NN1,NN2}. As we shall argue here large
$N_c$ QCD can provide additional insights into the nature of the
nucleon-nucleon force. In particular large $N_c$ QCD helps us to
understand both the nature of the meson-exchange picture of
nucleon-nucleon interactions and the limitations of such a picture. 
The large $N_c$ perspective also sheds light on the role of the
$\Delta$ resonance in nucleon-nucleon interactions. At a practical
level, knowledge of the special role played by  $\Delta$ in
canceling certain large contributions may prove to be useful in
constructing nucleon-nucleon interactions.

\begin{figure}[ht]
\begin{center}
\epsfig{figure=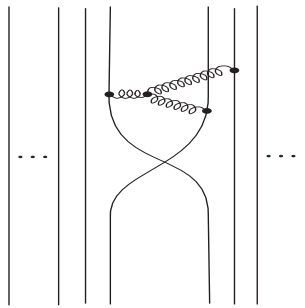,height=4cm,width=4cm,clip=,angle=90}
\bigskip
\caption{A typical diagram of order $N_c$ contributing to the nucleon-nucleon
scattering in the large $N_c$ limit.}
\label{fig1}
\end{center}
\end{figure}

\begin{figure}[t]
\begin{center}
\epsfig{figure=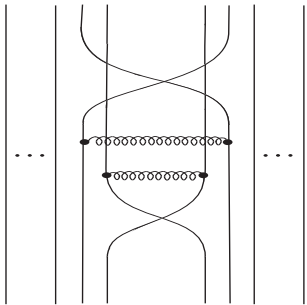,height=4cm,width=4cm,clip=,angle=90}
\bigskip
\caption{A typical diagram of order $N_{c}^{2}$.}
\label{fig2}
\end{center}
\end{figure}

\begin{figure}[h]
\begin{center}
\epsfig{figure=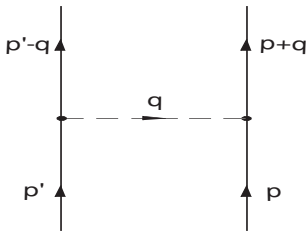,height=4cm,width=3cm,clip=,angle=90}
\bigskip
\caption{A one-meson exchange diagram contributing to the nucleon-nucleon
potential; initial and final nucleons are on shell; 
$p=(|\vec{p}|^2/2 m_N,\,-\vec{p})$,
$p^\prime=(|\vec{p}|^2/2 m_N,\,\vec{p})$ and
$q=(q^0,\,\vec{q})$ are energy-momentum 4-vectors flowing through the various
lines.}
\label{fig3}
\end{center}
\end{figure}

\begin{figure}[t]
\begin{center}
\epsfig{figure=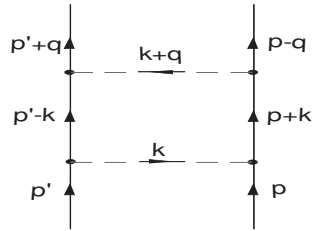,height=4cm,width=3cm,clip=,angle=90}
\bigskip
\caption{A box diagram; $p=(|\vec{p}|^2/2 m_N,\,-\vec{p})$,
$p^\prime=(|\vec{p}|^2/2 m_N,\,\vec{p})$, $k=(k^0,\,\vec{k})$ and 
$q=(q^0,\,\vec{q})$  are energy-momentum 4-vectors.}
\label{fig4}
\end{center}
\end{figure}

The first treatment of nucleon-nucleon interactions in large $N_c$
QCD was done by Witten in his seminal paper on baryons in the large
$N_c$ limit \cite{LN2}.  He argued that the dominant interaction
between two baryons is generically of order $N_c$.  His argument was
based on consideration of diagrams such as the one in
Fig.~\ref{fig1}. It is clear that such a diagram is order
$N_c$---a factor $N_{c}^3$ from combinatorics and a factor of
$N_{c}^{-2}$ from the coupling constants. It is straightforward to see
that all quark-line-connected graphs beginning and ending with two flavor
singlet combinations of $N_c$ quarks will be ${\cal O}(N_c)$ or
less. However, this interaction strength of order
$N_c$ cannot represent the strength of the nucleon-nucleon
scattering amplitude. In the first place, unitarity implies that the 
scattering amplitude does not grow without a bound as $N_c$ goes to infinity. 
Secondly, one can consider graphs like Fig.~\ref{fig2} which
although disconnected at the quark level, contribute at the
nucleon-nucleon level to the full interaction. It is
straightforward to see that the graph in Fig.~\ref{fig2} is order $N_c^2$
so that the interaction cannot simply go as $N_c$. One natural way
to interpret the physics contained in the diagrams typified by
Figs.~\ref{fig1} and \ref{fig2} is to argue that they get translated
at the hadronic level to contributions from nucleon-meson diagrams as
in Figs.~\ref{fig3} and \ref{fig4}. In such a hadron-based picture the
${\cal O}(N_{c}^{2})$ contribution then appears as the iteration of an
underlying ${\cal O}(N_c)$ interaction. One key issue addressed in
this paper concerns the nature of the translation from the
quark-gluon based diagrams of Figs.~\ref{fig1} and \ref{fig2} to the
hadronic based ones of Figs.~\ref{fig3} and \ref{fig4}.

Witten \cite{LN2} noted an additional difficulty of having 
nucleon-nucleon interaction scaling as $N_c$, there is no
description of the scattering process which possesses a smooth large
$N_c$ limit if the momenta are of order unity. The basic difficulty
in this case is that the kinetic energy of the nucleons is
generically much smaller than the potential energy and the interplay
of kinetic and potential energy which is at the crux of scattering cannot be
independent of $N_c$. Witten noted that if one works in a kinematic
regime with momenta of order $N_c$ ($\it i.e.$ an approach to the large $N_c$
limit with the nucleon velocities rather than momenta fixed), then
the kinetic and potential terms are of the same order so that a smooth
limit is possible. For this kinematic regime Witten suggested that
the scattering process can be described using the time-dependent
Hartree (TDH) approximation.  It is straightforward to see that the
TDH equations with fixed initial velocity have solutions which are
independent of $N_c$ . In practice, such TDH calculations have not
been done in QCD and would be very difficult for systems with light
quarks.

Here we wish to focus on a different limit than Witten's, {\it i.e.}
on low momentum nuclear reactions.  Accordingly we do not wish to
let the nucleon momenta scale with $N_c$; rather we will restrict our
attention to the kinematic regime of nucleon momenta of order
$N_{c}^{0}$. As noted by Witten, in such a regime there is no
smooth expression for the scattering amplitude.  However, as argued
by Kaplan and Savage \cite{NN1} and Kaplan and Manohar \cite{NN2} in
this regime one may identify the nucleon-nucleon interaction from
the quark line connected pieces as a nonrelativistic  potential
which has a dominant contribution of order $N_c$.  Such a description can be
interpreted on the hadronic level as a meson exchange. The one-meson exchange
potential (Fig.~\ref{fig3}) can be as large as ${\cal O} (N_c)$ since a
generic baryon-meson coupling is of order $\sqrt{N_c}$ \cite{LN2} and hence
the large $N_c$ scaling at the hadronic level is consistent with the
quark-gluon level. The key insight of Refs.~\cite{SF1,SF2,SF3} is that large
$N_c$ QCD implies an approximate contracted $SU(4)$ spin-isospin symmetry on
the baryons and that this symmetry imposes constraints on the dominant parts
of the potential. Thus, the dominant part of the nucleon-nucleon
interaction are constrained to be contracted $SU(4)$ symmetric and
terms which break this symmetry are suppressed by two powers of $N_c$. For
example, the dominant $({\cal O}(N_c))$ contribution to the tensor
force is proportional to $\vec{\tau}_1 \cdot \vec{\tau}_2$, while the isospin
independent part only contributes at order $N_{c}^{-1}$.

This paper will address a number of issues connected with the nucleon-nucleon
interaction in large $N_c$ QCD. One central issue is the identification of the
connected diagrams, such as Fig.~\ref{fig3}, as a nonrelativistic potential
(of order $N_c$), as was done in Refs.~\cite{NN1,NN2}. The argument for doing
this is clearly heuristic and is based on the notion that large interactions
must be iterated to all orders. Of course, a potential used in a Schr\"odinger
equation is iterated to all orders. This interpretation seems to resolve in a
simple manner the order $N_c^2$ contributions of Fig.~\ref{fig4}; it is just
one iteration of the potential (among an infinite number of possible
iterations).

Unfortunately, this heuristic argument is not unique: an alternative argument
would be to identify such an interaction with a  kernel of a Bethe-Salpeter
equation (with a strength of order $N_c$)  which again is to be
iterated to all orders.  Moreover, it is by no means clear that
these are equivalent---a Bethe-Salpeter kernel of order $N_c$ does not
necessarily imply that the potential in a nonrelativistic reduction
of the Bethe-Salpeter equation is order $N_c$. Thus, at the fundamental level
it has to be established whether large $N_c$  QCD is consistent with a
nonrelativistic nucleon-nucleon potential of order $N_c$ or with a
Bethe-Salpeter kernel of order $N_c$.  

A second fundamental issue in this paper is the extent to which large $N_c$
QCD justifies a meson exchange picture of nucleon-nucleon interactions. A
meson exchange is a natural way to understand nucleon-nucleon interactions as
arising from QCD: QCD leads to the existence of colorless hadronic
states---baryons and mesons---and the interactions between baryons
arise from the exchange of virtual mesons. Indeed, some
phenomenologically successful nucleon-nucleon potentials are based
directly on a meson exchange picture \cite{MNN}. On the other hand,
the argument for meson exchange dominating the nucleon-nucleon
interaction is not compelling and many equally successful
nucleon-nucleon potentials include only one-pion exchange treating
all shorter distance effects purely phenomenologically \cite{PhNN}. 
At first sight it might seem that large $N_c$ arguments do not
support the meson-exchange picture of nucleon-nucleon interactions. 
In the first place, as noted by Witten \cite{LN2} baryons in the
large $N_c$ behave as solitons, and when two solitons are brought
close enough to interact each one distorts in the presence of the
other yielding effects which cannot be easily described in
terms of meson exchange. Indeed Witten's prescription for
scattering for momenta of order $N_c$, TDH, necessarily builds in
these non-meson-exchange type effects; the clusters of $N_c$ quarks
which interact in TDH are {\it not} simply the Hartree
wave-functions for two nucleons.

\begin{figure}[h]
\begin{center}
\epsfig{figure=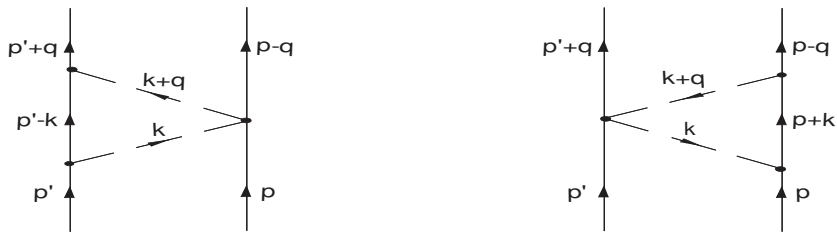,height=11cm,width=3cm,clip=,angle=90}
\bigskip
\caption{Triangle diagrams; $p=(|\vec{p}|^2/2 m_N,\,-\vec{p})$,
$p^\prime=(|\vec{p}|^2/2 m_N,\,\vec{p})$, $k=(k^0,\,\vec{k})$ and 
$q=(q^0,\,\vec{q})$ are energy-momentum 4-vectors.}
\label{fig6}
\end{center}
\end{figure}

\begin{figure}[h]
\begin{center}
\epsfig{figure=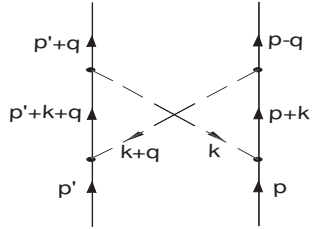,height=4cm,width=3cm,clip=,angle=90}
\bigskip
\caption{A crossed-box diagram; 
$p=(|\vec{p}|^2/2 m_N,\,-\vec{p})$,
$p^\prime=(|\vec{p}|^2/2 m_N,\,\vec{p})$, $k=(k^0,\,\vec{k})$ and
$q=(q^0,\,\vec{q})$ are energy-momentum 4-vectors.}
\label{fig5}
\end{center}
\end{figure}

There is a second reason why one might suspect that large $N_c$ QCD
does not justify a meson exchange point of view for nucleon-nucleon
interactions. The meson-exchange picture does not imply only single
meson exchanges but two or more meson exchanges as well.  Consider,
the large $N_c$ scaling of a generic two-meson exchange process.
Some typical diagrams contributing to the potential are shown in 
Fig.~\ref{fig6}, in which two exchange mesons are coupled to one of
the nucleons at a single vertex. Such diagrams are generically of
order $N_c$ since the coupling of the meson current to a nucleon is
of order $N_{c}^{0}$ \cite{LN2}; the additional power of $N_c$
comes from two nucleon-meson vertices (each of which contributes
$N_c^{1/2}$. Thus, these diagrams are consistent with the previously
deduced large $N_c$ scaling behavior of the potential.  However, if
one considers a generic crossed-box diagram as shown in Fig.~\ref{fig5} ,
one encounters an inconsistency. Since, the nucleon-meson coupling is
generically of order $\sqrt{N_c}$, the diagram in Fig.~\ref{fig5} is of order
$N_{c}^{2}$. This scaling, however, violates the proposed large
$N_c$ scaling of the nucleon-nucleon potential (which is supposed to
go as $N_c$).  Clearly, three and more meson exchange diagrams will
yield ever-larger inconsistencies.

This paper will also address another central issue in nucleon-nucleon physics;
namely, the role of the $\Delta$ resonance in intermediate states. As is well
known, the $\Delta$ resonance has a low excitation energy for large $N_c$
(with $m_\Delta - m_N \sim N_c^{-1}$) \cite{SF2,SK}. Moreover, the inclusion of
virtual $\Delta$'s is known to be essential to ensure the consistency of
large $N_c$ predictions for hadronic processes and its inclusion is
necessary for the contracted $SU(4)$ spin-flavor structure of baryons to be
manifest. Thus, one might expect that virtual $\Delta$'s will also play a key
role in nucleon-nucleon reactions in the large $N_c$ limit.

In this paper we will show the following picture is consistent with
large $N_c$ QCD:\\ 
i) The nucleon-nucleon interaction can be described by a potential of order
$N_c$ but {\it not} by a Bethe-Salpeter kernel of order $N_c$.  \\
ii) The meson-exchange picture can be used consistently to describe
nucleon-nucleon interactions for momenta of order $N_{c}^{0}$. \\
iii) The meson-exchange picture  of the nucleon-nucleon potential (of order
$N_c$) breaks down for momenta of order $N_c$.\\  
iv) The leading order nucleon-nucleon potential is symmetric under contracted
$SU(4)$ with corrections down by two powers in $N_c$ yielding the spin-flavor
structure of Ref.~\cite{NN1,NN2}. \\
v) The contracted $SU(4)$ structure implies
a central role for intermediate states containing $\Delta$'s.  

The basic strategy of this paper is to assume that the picture
outlined above is correct and then to  show that it does not lead to
any inconsistencies up to and including two-meson exchange
potentials.  The key difficulty which must be addressed is the
problem of the crossed-box graphs mentioned above: If point ii) is
correct then they must be included; however, they generically
contribute to the potential at order $N_c^2$ which exceeds the
potential's supposed order $N_c$ scale from point i). More generally
they lead to contributions that are inconsistent with the
spin-flavor structure of Refs.~\cite{NN1,NN2}.  However, as we will show, all
contributions from the crossed-box graphs which are inconsistent
with the spin-flavor structure of Refs.~\cite{NN1,NN2} are canceled by
contributions coming from the retardation effects in the box graphs.
While it has long been known that such a cancelation occurs for
scalar isoscalar mesons between retardation effects in the box graph
and the crossed-box graph \cite{cancel}, it has not been previously
shown that such cancelations are far more general and protect the
large $N_c$ structure of the nucleon-nucleon potential including the
hierarchy of large and small contributions in terms of spin and
isospin.  It will also be shown that such a cancelation does not
occur if $\Delta$ intermediate states are excluded or if the momenta
of the nucleons is of order $N_c$.

\section{Review of the $SU(4)$ contracted symmetry}

The baryon sector of large $N_c$ QCD exhibits an approximate $SU(4)$ 
contracted light quark spin-flavor symmetry \cite{SF1,SF2,SF3}.
The contracted $SU(4)$ algebra is:
\begin{eqnarray} 
&\left [J^i \, , J^j \right]=i \, \epsilon_{ijk} \, J^k \, ,
\,\,\,\, \,\,\,\,\,\,\,\,
\left [I^a \, , I^b \right]  =  i\, \epsilon_{abc} \, I^c \, , 
\,\,\,\, \,\,\,\,\,\,\,\,\,
\left [J^i \, , I^a \right]=0 \, , 
\,\,\,\,\,\,\,\,\,\,\,\,\,\, \,\,\,\,\,\,\,\,\,\,\,\,\,\,\,\,&
\nonumber \\
& \left [J^i \, , X_{0}^{jb} \right]=i \, \epsilon_{ijk} \, X_{0}^{kb} \, , 
\,\,\,\,\,\,\,\,\,\,\,\,\,\,
\left [I^a \, , X_{0}^{jb} \right]=i\, \epsilon_{abc} \, X_{0}^{jc} \, , 
\,\,\,\, \,\,\,\,\,\,\,\,
\left [X_{0}^{ia} \, , X_{0}^{jb} \right]=0 \, , &
\label{SU4}
\end{eqnarray}	
where $i, a=1,2,3$ are spin and isospin indices.

\begin{table}[t]
\caption{Nonrelativistic scalar and pseudo-scalar meson-baryon couplings. 
The ground state baryons, B, belong to an irreducible representation
$I=J$.}
\begin{tabular}{|r|r|r|r|r|}
& \multicolumn{2}{c|}{Scalars} & \multicolumn{2}{c|}{Pseudo-scalars} \\
& $I=0$ & $I=1$ & $I=0$ & $I=1$ \\
\hline
meson & $f_0$ & $a_0$ & $\eta$  & $\pi$ \\
\hline
meson-baryon coupling & $ B^\dag B \phi $ & $ B^\dag I^a B \phi^a $ &
$ B^\dag J^i B \partial^i \phi $ & 
$ B^\dag X^{ia} B \partial^i \phi^a $ \\
\hline
scaling of the coupling & $\sqrt{N_c}$ & $(\sqrt{N_c})^{-1}$ &
$(\sqrt{N_c})^{-1}$ & $\sqrt{N_c}$ \\
\hline
spin-flavor term & $ V_{0}^{0} $ & $ V_{0}^{1} $ & $ V_{T}^{0} $ & 
$ V_{T}^{1} $ \\
\hline
KSM scaling & $N_c$ & $N_{c}^{-1}$ & $N_{c}^{-1}$ & $N_c$
\end{tabular}
\label{tab1}
\end{table}

\begin{table}[ht]
\caption{Nonrelativistic vectors and pseudo-vector meson-baryon couplings.
The ground state baryons, B, belong to an irreducible representation
$I=J$.}
\begin{tabular}{|r|r|r|r|r|r|r|}
& \multicolumn{4}{c|}{Vectors} & \multicolumn{2}{c|}{Pseudo-vectors} \\
& \multicolumn{2}{c|}{$I=0$} & \multicolumn{2}{c|}{$I=1$}& $I=0$ & $I=1$ \\
\hline
meson & $\omega^t$ & $\vec{\omega}$ & $\rho^t$ & $\vec{\rho}$ & 
$\vec{f}_1$ & $\vec{a}_1$ \\
\hline
meson-baryon coupling &
$ B^\dag B V^t $ & $ B^\dag \epsilon_{ijk} J^k B \partial^i V^j$ &
$ B^\dag I^a B  V^{ta} $ & 
$B^\dag \epsilon_{ijk} X^{ka} B \partial^i V^{ja}$ &
$ B^\dag J^i B  A^i $ & $ B^\dag X^{ia} B A^{ia} $ \\ 
\hline
scaling &
$\sqrt{N_c}$ & $(\sqrt{N_c})^{-1}$ & $(\sqrt{N_c})^{-1}$ & $\sqrt{N_c}$ & 
$(\sqrt{N_c})^{-1}$ & $\sqrt{N_c}$ \\
\hline
spin-flavor term &
$ V_{0}^{0} $ & $ V_{T}^{0} $ & $ V_{0}^{1} $ & $ V_{T}^{1} $ & 
$ V_{\sigma}^{0} $ & $ V_{\sigma}^{1} $ \\
\hline
KSM scaling &
$N_c$ & $N_{c}^{-1}$ & $N_{c}^{-1}$ & $N_c$ & 
$N_{c}^{-1}$ & $N_c$ 
\end{tabular}
\label{tab2}
\end{table}

In large $N_c$ two-flavor QCD, baryons belong to the infinite dimensional
irreducible representation of the contracted $SU(4)$ algebra with
$I=J=1/2, 3/2, 5/2, ...$ \cite{SF1,SF2,SF3}. For $N_c =3$ the $I=J=1/2$ and
$I=J=3/2$ states are identified with nucleon and $\Delta$. Other states are
presumably a large $N_c$ artifact. The meson-baryon couplings connecting the
states with different spin and isospin are given in terms of the matrix
elements of $X^{ia}$ which is defined by its matrix elements between baryon
states ({\it e.g.} Eq.~(\ref{CG})). This operator is equal to $X_{0}^{ia}$ at
leading order in the $1/N_c$ expansion. As shown in Ref.~\cite{SF2}, 
the next-to-leading order term is proportional to $X_{0}^{ia}$
\begin{equation}
X^{ia}=(1+{\alpha \over N_c}) X_{0}^{ia}+{\cal O}(1/N_{c}^{2}) \, ,
\label{X}
\end{equation}
where $\alpha$ is a constant independent of the spin and isospin indices. As a
result, the spin-flavor operators $X^{ia}$ commute up to
${\cal O}(N_{c}^{-2})$ corrections:
\begin{equation}
\left[ X^{ia} \, , \, X^{jb} \right]={\cal O}({1\over N_{c}^{2}}) \,.
\label{XX}
\end{equation}	

Meson-baryon couplings satisfying the contracted spin-flavor symmetry
are listed in the third row of  Table~\ref{tab1} and Table~\ref{tab2}.
These couplings are obtained from the nonrelativistic reduction of the
corresponding covariant Yukawa couplings with corrections suppressed by
$1/N_c$. The matrix elements of the $X_{0}^{ia}$ generators between the baryon
states are given in terms of the Clebsch-Gordan coefficients by \cite{SF2}:
\begin{equation}
\langle I^\prime \, I_{3}^\prime \, , J^\prime \, J_{3}^\prime |
X_{0}^{ia} | I \, I_{3} \, , J \, J_{3} \rangle =
\sqrt{(2J+1) \over (2 J^\prime +1)} \,
\langle J^\prime\,, J_{3}^{\prime} | J\, J_{3}\, ; 1\, i \rangle
\,
\langle I^\prime\,, I_{3}^{\prime} | I\, I_{3}\, ; 1\, a \rangle
\,,
\label{CG}
\end{equation}
where only spin and isospin labels of the baryon states are shown explicitly.

We will need to know the matrix elements of the anticommutators between the
nucleon states. When restricting attention to nucleon initial and final states
one can easily deduce:
\begin{equation}
\left\{J^{i}\,,\, X^{ja}\right\}_{N^\prime N}= 
{1\over 2}\,\delta^{ij}\, I^a +{\cal O}({1\over N_{c}^{2}})\, ,
\,\,\,\,\,\,\,
\left\{I^{a}\,,\, X^{ib}\right\}_{N^\prime N}= {1\over 2}\,\delta^{ab} \,
J^i + {\cal O}({1\over N_{c}^{2}}) \, ,
\label{anticom}
\end{equation}
where we have used the fact that $J^i$ and $I^a$ only take nucleons
into nucleons. The ${\cal O} (1/N_c)$ corrections in Eq.~(\ref{anticom})
vanish due to the fact that the constant $\alpha$ in Eq.(\ref{X}) is
independent of the spin and isospin.

The large $N_c$ scaling of the baryon matrix elements have been analyzed
in Refs.~\cite{SF2,SF3}. Since a general one-quark operator ({\it e.g.} 
axial vector current) can couple to any of the $N_c$ quarks in a baryon, its
matrix elements between ground state baryons are of order $N_c$ (providing
the cancelation between different quark line insertions does not occur).
The operators with spin-flavor structure given by ${\bf 1}$ and $X^{ia}$
behave in this leading fashion \cite{SF2}. On the other hand, currents
containing only $J^i$ and $I^a$ are of order $N_{c}^{0}$. Heuristically, the
reason is that for baryons with $J=I=1/2, 3/2, ...$ only one out of $N_c$
quarks carry the spin and isospin quantum numbers of the state. The large
$N_c$ scaling of a meson-baryon coupling is obtained by dividing the
corresponding current matrix element by the meson decay constant which is of
order $N_{c}^{-1/2}$. Hence, the meson-baryon couplings containing spin-flavor
operators ${\bf 1}$ or $X^{ia}$ are of order  $N_{c}^{1/2}$. Examples of
such leading couplings are the couplings of $f_0$ and $\pi$ mesons to baryons.
In addition, the time component of $\omega$ ($\omega^t$) and spatial
components of $\rho$ $(\vec{\rho})$ and $a_1$ $(\vec{a}_1)$ couple to the
baryons with a strength proportional $N_{c}^{1/2}$. Couplings containing
$J^i$ and $I^a$ are of order $N_{c}^{-1/2}$. The examples include the
couplings of $a_0$ and $\eta$, spatial components of $\omega$ and $f_1$
($\vec{\omega}$ and $\vec{f}_1$) and the time component of $\rho$ $(\rho^t)$.
These counting rules are listed in the fourth row of Table~\ref{tab1} and
Table~\ref{tab2}. 

Similarly, the spin-flavor structure of the nonrelativistic nucleon-nucleon 
potential can be analyzed in the large $N_c$ QCD. The general form of this
potential is:
\begin{eqnarray}
& V_{NN}=V_{0}^{0}+V_{\sigma}^{0}\,\vec{\sigma}_{(1)} \cdot\vec{\sigma}_{(2)}
+V_{T}^{0} S_{12} +V_{LS}^{0} \, \vec{L}\cdot \vec{S} +V_{Q}^{0}\, Q_{12}&
\nonumber \\
& +\left(
V_{0}^{1}+V_{\sigma}^{1}\,\vec{\sigma}_{(1)} \cdot\vec{\sigma}_{(2)}
+V_{T}^{1} S_{12} +V_{LS}^{1} \, \vec{L}\cdot \vec{S} +V_{Q}^{1}\, Q_{12}
\right) \vec{\tau}_{(1)}\cdot\vec{\tau}_{(2)}\,,&
\label{VNN}
\end{eqnarray}
where
\begin{equation}
S_{12}=3 \, \vec{\sigma}_{(1)} \cdot \hat{r} \, 
\vec{\sigma}_{(2)} \cdot \hat{r} - \vec{\sigma}_{(1)}\cdot\vec{\sigma}_{(2)}
\,, \,\,\,\,\,\,\,\,\,\,\,\,\,\,\,\,\,\,\,\,\,
Q_{12}={1\over2} \left\{\vec{\sigma}_{(1)}\cdot \vec{L}\,,\,
\vec{\sigma}_{(2)}\cdot \vec{L} \right\} \,,
\label{SQ}
\end{equation}
where $\vec L$ and $\vec S$ are the total orbital and spin angular momenta
of the system of two nucleons.  
The operators in Eq.~(\ref{VNN}) multiplying the position and velocity 
dependent functions $V_{n}^{1,2}$ ($n=0,\,\sigma,\,T,\,LS,\,Q$) are referred 
to as the central, spin-spin, tensor, spin-orbit and quadratic spin-orbit 
components of the nucleon-nucleon potential in the isosinglet and isotriplet 
channels. 

In Refs.~\cite{NN1,NN2}, the large $N_c$ scaling of functions $V_{n}^{1,2}$ was
analyzed using the spin-flavor counting rules of the generators of the 
contracted $SU(4)$. The analysis is based on two assumptions. One is that
the nucleon-nucleon interaction can be described by an Hartree Hamiltonian
which can be written as a sum of operators with particular 
spin-flavor structure satisfying the large $N_c$ scaling rules of the 
contracted $SU(4)$ symmetry. In addition, the authors implicitly assumed that
the Hartree picture leads to a potential of order $N_c$ for momenta of order
one.
At the hadronic level, the latter assumption is essentially equivalent to a 
one-meson exchange picture of the potential. Based on the above assumptions,
the following counting rules were obtained in Refs.~\cite{NN1,NN2}:
\begin{equation}
V_{0}^{0} \sim V_{\sigma}^{1} \sim V_{T}^{1} \sim N_c \, ,
\,\,\,\,\,\,\,\,\,\,\,\,\,\,\,\,\,\,\,\,\,\,\,\,\,\,\,\,
V_{0}^{1} \sim V_{\sigma}^{0} \sim V_{T}^{0} \sim {1\over N_c} \,. 
\label{KSM}
\end{equation}
In addition, the spin-orbit and quadratic spin-orbit suppressed by 
$1/m_{B}\sim 1/N_c$ ($m_{B}$ is a baryon mass) are of order $1/N_{c}^{2}$.
The scaling rules in Eq.~(\ref{KSM}) will be referred to as KSM counting rules.

It is easy to see how the counting rules in Eq.~(\ref{KSM}) arise from the
large $N_c$ scaling of the meson-baryon couplings at the one-meson
exchange level. At this level, a given term in the potential, Eq.~(\ref{VNN}),
scales as the square of the corresponding coupling constant. Since, for
example, the isoscalar central
potential at leading order gets contributions from $f_0$ exchange it is of
order $(\sqrt{N_c})^2 = N_c$. Similarly, the one-pion exchange contributes
to the leading part of the isovector tensor term which, therefore, scales as
$(\sqrt{N_c})^2 = N_c$. On the other hand, the isoscalar tensor potential,
$V_{T}^{0}$, is of order  $N_{c}^{-1}$ since its leading contribution is
from one-$\eta$ exchange. The leading contributions at the one-meson exchange
level
are shown in the fifth row of Table~\ref{tab1} and Table~\ref{tab2}; the $N_c$ 
scaling of these contributions are shown in the last row of Table~\ref{tab1} 
and Table~\ref{tab2}. We will show that the nucleon-nucleon potential is 
consistent with KSM counting rules, Eq.~(\ref{KSM}), up to and including
two-meson exchange contributions.

\section{Two-meson exchange contributions}

The Feynmann diagrams contributing at the two-meson exchange level are shown
in Figs.~\ref{fig4},~\ref{fig5}, and \ref{fig6}---the box, the crossed-box and
the triangle graphs. In these diagrams the initial
and final nucleons are on-mass shell. This condition is necessary if the 
diagrams are used to derive the nucleon-nucleon potential. The baryon 
energy-momentum relation is treated nonrelativistically with the baryon 
propagators having the following form:
\begin{equation}
{i\over k^0-|\vec k|^2/2 m_B +i \epsilon} \left (1+{\cal O}({1\over N_c})
\right)\, ,
\label{NRprop}
\end{equation}
where $k^0$ and $\vec{k}$ are the energy and the momentum of an intermediate 
baryon with mass $m_B$. In practice, $m_B=m_N+{\cal O}(1/N_c)$ 
($m_N$ is the nucleon mass) for the ground state baryons with $I=J=1/2, 3/2,
5/2$, etc. Relativistic effects are suppressed by $1/m_B \sim N_{c}^{-1}$.
The mesons are treated in a fully relativistic form. The meson-baryon vertices
are in general momentum and energy dependent. Note, the time and spatial
components of $\omega$ and $\rho$ have different couplings at leading
nonrelativistic order (Table~\ref{tab1} and Table~\ref{tab2}). In addition,
the spin-flavor structure of $\omega^t$ coupling 
is identical to that of $f_0$ (or $\sigma$). Similarly, $\rho^t$ and $a_0$ 
couplings have identical spin-flavor structure.

A two-meson exchange diagram may contain a piece which is equal to the one 
iteration of the potential. These contributions will be included when solving 
the Shr\"odinger equation and must be excluded from the 
nucleon-nucleon potential to avoid double counting. This can be illustrated 
using the the two-scalar exchange diagrams. 

The contribution to the nucleon-nucleon potential from a one-scalar 
exchange, Fig.~\ref{fig3}, with point couplings is given by,
\begin{equation}
V_{f_0} (\vec q)= {g_{f_0}^{2} \over (q^{0})^{2}-|\vec{q}|^2-m_{f_0}^{2}}
={-g_{f_0}^{2} \over |\vec{q}|^2+m_{f_0}^{2}}
\left(1+{\cal O}(1/N_{c}^{2})\right) \,
\label{Vf0}
\end{equation}
where $m_{f_0}$ (${\cal O}(N_{c}^{0})$) is the mass of the $f_0$ meson, and
the coupling constant $g_{f_0}$ is of order $\sqrt{N_c}$ (Table~\ref{tab1}).
Note that $(q^{0})^{2}$ can be neglected since $q^0$ is of order 
$N_{c}^{-1}$. 

Similarly, the contribution of the two-scalar exchange box diagram, 
Fig.~\ref{fig4} to the scattering amplitude, ${\cal M}$, is 
\begin{eqnarray}
& i\, {\cal M}_{\Box}= \int {d^3 k\over (2\pi)^3} \int {d k^0 \over 2 \pi} 
{g_{f_0}^4 \over ((k^0)^2-|\vec{k}|^2-m_{f_0}^{2}) 
((k^0+q^0)^2-|\vec{k}+\vec{q}|^2-m_{f_0}^{2})} &
\nonumber \\ 
& \times {1\over (k^0 + {|\vec{p}|^2/ 2 m_B}-{|\vec{p}-\vec{k}|^2/ 2 m_B})
(-k^0 + {|\vec{p}|^2 / 2 m_B}-{|\vec{p}-\vec{k}|^2 / 2 m_B})} \, .&
\label{Mbox}
\end{eqnarray}
It is convenient to first perform the $k^0$ integral. There are two classes of
poles in the complex $k^0$ plane, namely from the baryon and meson 
propagators. It is easy to see that the baryon poles in Eq.~(\ref{Mbox}) are
on the opposite side of the real $k^0$ axis. By closing the integration
contour in the upper or lower complex plane only one of these baryon poles
will contribute to ${\cal M}_{\Box}$. Closing the contour in the upper plane
we get for the baryon pole contribution:
\begin{equation}
i\,{\cal M}_{\Box}^{B}=\int {d^3 k\over (2\pi)^3}  {i \, g_{f_0}^{4} 
\over (|\vec{k}|^2+m_{f_0}^{2})(|\vec{k}+\vec{q}|^2+m_{f_0}^{2}) 
(|\vec{p}|^2/  m_B-|\vec{p}-\vec{k}|^2 / m_B)}
\left(1+{\cal O}({1\over N_c})\right) \, ,
\label{MB}
\end{equation}
where, in addition to $q^0 \sim N_{c}^{-1}$, the position
of the baryon pole, $k^{0}_{B}=(|\vec{p}|^2-|\vec{p}-\vec{k}|^2)/2 m_{B}$, is
neglected when the meson propagators are evaluated. However, the position of 
the baryon pole is of leading order when the other baryon propagator is 
evaluated. As will become clear, this value of a baryon propagator is 
identical to the nonrelativistic Green function in the Lippman-Schwinger
equation for the scattering amplitude.

The baryon pole contribution, ${\cal M}_{\Box}^{B}$, should be compared with
one iterate of the Lippman-Schwinger equation (in the center-of-mass frame):   
\begin{equation}
T(\vec{p}, \vec{p}+\vec{q})= -\, V(\vec{q})+\int {d^3 k\over (2\pi)^3}
V(-\vec{k}) \, G_0 (\vec{k}) \, T(\vec{p}-\vec{k}, \vec{p}+\vec{q})\, ,
\label{LpSw}
\end{equation}
where the nonrelativistic baryon Green function is given by
\begin{equation}
G_0 (\vec{k}) \equiv 
{1\over(|\vec{p}-\vec{k}|^2-|\vec{p}|^2)/m_B + i \epsilon} \, .
\label{G0}
\end{equation}
The first term in Eq.~(\ref{LpSw}) corresponds to a potential at the one-meson 
exchange level. For the one-scalar exchange this potential is given in
Eq.~(\ref{Vf0}). Iterations of the Lippman-Schwinger equation lead to: 
\begin{eqnarray}
&T(\vec{p}, \vec{p}+\vec{q})=-\, V(\vec{q})+\int {d^3 k\over (2\pi)^3}
V(-\vec{k}) \, G_0 (\vec{k})\, V(\vec{k}+\vec{q})&
\nonumber \\
& - \int {d^3 k\over (2\pi)^3}\int {d^3 k^\prime \over (2\pi)^3}
V(-\vec{k}) \, G_0 (\vec{k})\, V(\vec{k}+\vec{k}^\prime) G_{0}(\vec{k}^\prime)
V(\vec{q}-\vec{k}^\prime) + ... \, ,&
\label{itt}
\end{eqnarray}
where the ellipsis indicates higher-order iterations. For the two-scalar
exchange, the first iteration of the potential in Eq.~(\ref{Vf0}) is:
\begin{equation}
\int {d^3 k\over (2\pi)^3} V_{f_0} (-\vec{k}) \, G_0 (\vec{k})\, 
V_{f_0}(\vec{k}+\vec{q})
=\int {d^3 k\over (2\pi)^3} \, {g_{f_0}^{4} 
\over (|\vec{k}|^2+m_{f_0}^{2})(|\vec{k}+\vec{q}|^2+m_{f_0}^{2}) 
(|\vec{p}-\vec{k}|^2 / m_B-|\vec{p}|^2/  m_B)} \, ,
\label{itt1}
\end{equation}
which is exactly equal to the baryon pole contribution, ${\cal M}_{\Box}^{B}$
(Eq.~(\ref{MB})), evaluated with the nonrelativistic baryon propagator,
Eq.~(\ref{NRprop}). Thus, the baryon pole contribution of the two-scalar box
diagram should not be included in the nucleon-nucleon potential. Note, that
the equality holds only if nonrelativistic baryon propagator is used to
evaluate ${\cal M}_{\Box}^{B}$.

The remaining contribution to ${\cal M}_{\Box}$ is from the meson poles. This
contribution is often referred to as the 
retardation effect since it is absent when using a static potential. The
retardation effect for two-scalar exchange is of order $N_{c}^{2}$
(see Table~\ref{tab1}), {\it i.e} it is larger than allowed by KSM
counting rules, Eq.~(\ref{KSM}). Hence, for the two-scalar
exchange diagrams to be consistent with the counting rules, the retardation
effect has to be cancelled by the crossed-box diagram. The key issue is whether
this cancelation indeed happens.

The baryon pole contribution in the box diagram has been discussed for the 
two-scalar exchange with point couplings. In fact, it can easily be
generalized for any two-meson exchange with general vertex functions. Indeed,
the above proof that the baryon pole contribution to the box diagram is one
iterate of the potential rests only on the nonrelativistic form of the
two-baryon propagators and the direction of the loop momenta and energy flow
through the baryon lines. Neither the spin-flavor structure nor the vertex
functions can change the position of the baryon poles. Thus, the baryon poles
from any two-meson box diagrams do not contribute to the nucleon-nucleon
potential. 

We have shown so far that the baryon pole contributions from the two-meson box
diagrams should not be included in the nucleon-nucleon potential. However,
the retardation effect and the crossed-box contribution can each be larger
than allowed by
the KSM counting rules. For example, the retardation effect and crossed-box
diagrams corresponding to the two-pion exchange are each of order $N_{c}^{2}$.
Moreover, the two-meson exchange diagrams in general can contribute to
different spin-flavor structures in the nucleon-nucleon potential,
Eq.~(\ref{VNN}). As a result, these contributions considered separately may
violate the KSM counting rules of the subleading $({\cal O}(N_{c}^{-1}))$ terms
in the potential. For example, two-pion exchange box and crossed-box diagrams
(each of order $N_{c}^{2}$) contribute not only to $V_{T}^{1}$ but among
others to isosinglet tensor force, $V_{T}^{0}$, as well. The latter, however,
should be of order $N_{c}^{-1}$ according to Eq.~(\ref{KSM}). Fortunately,
as will be shown below, the retardation effects cancel against the crossed-box
diagram contributions in all such cases.

A cancelation between the retardation effect and the crossed-box is well known
for the two-scalar exchange diagrams \cite{cancel}. The meson pole
contribution to ${\cal M}_{\Box}$, Eq.~(\ref{Mbox}), is: 
\begin{eqnarray}
&{\cal M}_{\Box}^{ret}=
\int {d^3 k\over (2\pi)^3} \int {d k^0 \over 2 \pi} 
2\, Im \left[{g_{f_0}^{4} 
\over ((k^0)^2-|\vec{k}|^2-m_{f_0}^{2}) 
((k^0+q^0)^2-|\vec{k}+\vec{q}|^2-m_{f_0}^{2})}\right] &
\nonumber \\
&\times 
P\left[{1\over (k^0 + {|\vec{p}|^2/ 2 m_B} - {|\vec{p}-\vec{k}|^2/ 2 m_B})
(-k^0 + {|\vec{p}|^2 / 2 m_B}-{|\vec{p}-\vec{k}|^2 / 2 m_B})}\right] \, &
\nonumber \\
& =\int {d^3 k\over (2\pi)^3} {g_{f_0}^{4} 
\over |\vec{k}+\vec{q}|^2-|\vec{k}|^2}
\left({1\over 2 (|\vec{k}|^2 +m_{f_0}^{2})^{3/2}}-
{1\over 2 (|\vec{k}+\vec{q}|^2 +m_{f_0}^{2})^{3/2}}\right)
\left(1+{\cal O}({1\over N_c})\right) \, ,&
\label{MBm}
\end{eqnarray}
where in the second step we have again neglected terms in the denominators
suppressed by $1/m_{B}\sim 1/N_{c}$ including the energy $q^0$; a symbol
$P$ indicates a principle value.

The contribution to the scattering amplitude from the two-meson crossed-box 
diagram, Fig.~\ref{fig5}, is   
\begin{eqnarray}
& i\,{\cal M}_{X}= \int {d^3 k\over (2\pi)^3} \int {d k^0 \over 2 \pi} 
{g_{f_0}^{4} \over 
((k^0)^2-|\vec{k}|^2-m_{f_0}^{2}) 
((k^0+q^0)^2-|\vec{k}+\vec{q}|^2-m_{f_0}^{2})} &
\nonumber \\
&\times \, {1\over 
(k^0 + q^0 + {|\vec{p}|^2/ 2 m_B}-{|\vec{p}+\vec{k}+\vec{q}|^2/ 2 m_B})
(k^0 + {|\vec{p}|^2 / 2 m_B}-{|\vec{p}-\vec{k}|^2 / 2 m_B})} \,. &
\label{MX}
\end{eqnarray}
Note, the baryon poles are now on the same side of the real axis in the $k^0$
complex plane. Hence, they do not contribute to ${\cal M}_{X}$.
The only non-vanishing contribution is from the meson poles:
\begin{eqnarray}
& {\cal M}_{X}=
\int {d^3 k\over (2\pi)^3} \int {d k^0 \over 2 \pi} 
2\, Im\left[{g_{f_0}^{4} \over ((k^0)^2-|\vec{k}|^2-m_{f_0}^{2}) 
((k^0+q^0)^2-|\vec{k}+\vec{q}|^2-m_{f_0}^{2})}\right] &
\nonumber \\ 
& \times P\left[{1\over (k^0 + q^0 + {|\vec{p}|^2/ 2 m_B}-
{|\vec{p}+\vec{k}+\vec{q}|^2/ 2 m_B})
(k^0 + {|\vec{p}|^2 / 2 m_B}-{|\vec{p}-\vec{k}|^2 / 2 m_B})}\right] \, &
\nonumber \\
&= -  \int {d^3 k\over (2\pi)^3}  
{g_{f_0}^{4}  \over |\vec{k}+\vec{q}|^2-|\vec{k}|^2}
\left({1\over 2 (|\vec{k}|^2 +m_{f_0}^{2})^{3/2}}-
{1\over 2 (|\vec{k}+\vec{q}|^2 +m_{f_0}^{2})^{3/2}}\right)
\left(1+{\cal O}({1\over N_c})\right) \, ,&
\label{MXm}
\end{eqnarray}
where the same approximations as in Eq.~(\ref{MBm}) were made. As evident from
Eqs.~(\ref{MBm}) and (\ref{MXm}), the retardation effect and the the
crossed-box diagram contribution for the two-scalar exchange cancel out up to
corrections of order $N_{c}^{-1}$:
\begin{equation}
{\cal M}_{\Box}^{ret}+{\cal M}_{X}={\cal O}({1\over N_c}) \, .
\label{cancel}
\end{equation}

It is important to stress, however, that the above cancelation does not occur
when the nucleon momenta are of order $N_{c}$ since for the momenta of order
$N_{c}$ the baryon propagators evaluated at the meson poles are different
(as can be seen from Eqs.~(\ref{MBm}) and (\ref{MXm})). Consequently, for the
momenta of order $N_{c}$, the nucleon-nucleon interaction cannot be
interpreted as a simple meson exchange picture consistent with the KSM
counting rules, Eq.~(\ref{KSM}). As will be shown below, the cancelation in
Eq.~(\ref{cancel}) is far more general. In fact, it occurs for all two-meson
exchange graphs provided the nucleon momenta are of order $N_{c}^{0}$ and the
meson-baryon couplings are contracted $SU(4)$ symmetric. Let us consider a
general box and crossed-box diagram containing any pair of intermediate mesons.

We will use Greek symbols to indicate an exchanged meson,
{\it e.g.} $\alpha=f_0,\,\rho,\,\pi$, etc. A given graph contains four vertex
functions, one for each meson-baryon coupling. The product of these four
functions will be denoted by
$\tilde{V}_{\alpha\beta}(k^0, \vec{k}, q^0, \vec{q})$. The function
$\tilde{V}_{\alpha\beta}(k^0, \vec{k}, q^0, \vec{q})$ does not contain
spin-flavor matrices of the corresponding meson-baryon couplings which 
will be written explicitly. It is clear that
$\tilde{V}_{\alpha\beta}=\tilde{V}_{\beta\alpha}$. To simplify formulae we
combine the product of $\tilde{V}_{\alpha\beta}(k^0, \vec{k}, q^0, \vec{q})$
and two meson propagators into a single energy-momentum dependent function
$V_{\alpha\beta}(k^0, \vec{k}, q^0, \vec{q})$ defined by
\begin{equation}
V_{\alpha\beta}(k^0, \vec{k}, q^0, \vec{q}) \equiv
{\tilde{V}_{\alpha\beta}(k^0, \vec{k}, q^0, \vec{q})
\over ((k^0)^2-|\vec{k}|^2-m_{\alpha}^{2}) \, 
((k^0+q^0)^2-|\vec{k}+\vec{q}|^2-m_{\beta}^{2})}  \,,
\label{Valpha}
\end{equation}
where $m_{\alpha}$ and $m_{\beta}$ are the masses of the $\alpha$ and $\beta$
mesons. The above-defined function is symmetric under the interchange of the
exchanged mesons, $V_{\alpha\beta}=V_{\beta\alpha}$. The analytic structure of
$V_{\alpha\beta}(k^0, \vec{k}, q^0, \vec{q})$ as a function of a complex
variable $k^0$ determines the retardation effects of the box graphs and the
contribution of the crossed-box diagrams. 
 
The spin-flavor structure of the meson-baryon vertices will be denoted by 
$\Gamma^{A}_{\alpha (n)} (\vec k)$ where a superscript $A$ specifies the
spin-flavor indices and subscript $n=1,2$ indicates to which of the two baryon
lines a meson couples. The momentum dependence arises in the case of
derivatively coupled mesons such as pions. The product of the two
$\Gamma^{A}_{\alpha (n)} (\vec k)$ structures at two ends of the same meson
propagator are constrained by the spin and isospin of the exchanged meson. To
enforce these constraints in the product $\Gamma^{A}_{\alpha (1)} (\vec k) \,
\Gamma^{B}_{\alpha (2)} (-\vec k)$, we introduce a symbol $C^{\alpha}_{AB}$.
For example, in the case when the exchanged meson is a pion, the above
product takes the form:
\begin{equation}
C^{\pi}_{AB} \Gamma^{A}_{\pi (1)} (\vec{k})
\Gamma^{B}_{\pi (2)} (-\vec{k}) 
= - \, \delta^{ij}\,\delta^{mn}\, \delta^{ab}\, k^j \, k^n \,
X^{ia}_{(1)} X^{mb}_{(2)}= - \, k^i \, k^m \,X^{ia}_{(1)} X^{ma}_{(2)} \, \,
\label{pion}
\end{equation} 
where the contracted $SU(4)$ pion-baryon coupling is used.

Using the above notation a contribution from a general box graph,
Fig.~\ref{fig4}, has the following form:
\begin{equation}
i\,{\cal M}_{\Box}= g_{\alpha}^{2} \, g_{\beta}^{2}
\int {d^3 k\over (2\pi)^3} \int {d k^0 \over 2 \pi}
{V_{\alpha\beta}(k^0, \vec{k}, q^0, \vec{q})  
C^{\alpha}_{AB} C^{\beta}_{CD}
\Gamma^{A}_{\alpha (1)} (\vec{k})\Gamma^{C}_{\beta (1)} (-(\vec{k}+\vec{q}))
\Gamma^{B}_{\alpha (2)} (-\vec{k}) \Gamma^{D}_{\beta (2)} (\vec{k}+\vec{q}) 
\over (k^0 + {|\vec{p}|^2/ 2 m_B}-{|\vec{p}-\vec{k}|^2/ 2 m_B})
(-k^0 + {|\vec{p}|^2 / 2 m_B}-{|\vec{p}-\vec{k}|^2 / 2 m_B})} \, ,
\label{Mboxgen}
\end{equation}
where $g_{\alpha}$ and $g_{\beta}$ are the corresponding coupling constants and
the momenta directions are shown in Fig.~\ref{fig4}. The crossed-box,
Fig.~\ref{fig5}, has a similar expression with the last two $\Gamma$'s
interchanged:
\begin{equation}
i\,{\cal M}_{X}= g_{\alpha}^{2} \, g_{\beta}^{2}
\int {d^3 k\over (2\pi)^3} \int {d k^0 \over 2 \pi}
{V_{\alpha\beta}(k^0, \vec{k}, q^0, \vec{q})
C^{\alpha}_{AB} C^{\beta}_{CD}
\Gamma^{A}_{\alpha (1)} (\vec{k})\Gamma^{C}_{\beta (1)} (-(\vec{k}+\vec{q}))
\Gamma^{D}_{\beta (2)} (\vec{k}+\vec{q}) \Gamma^{B}_{\alpha (2)} (-\vec{k}) 
\over (k^0 + q^0 + {|\vec{p}|^2/ 2 m_B}-{|\vec{p}+\vec{k}+\vec{q}|^2/ 2 m_B})
(k^0 + {|\vec{p}|^2 / 2 m_B}-{|\vec{p}-\vec{k}|^2 / 2 m_B})} \,. 
\label{MXgen}
\end{equation}
Note the difference in the baryon propagators relative to Eq.~(\ref{Mboxgen})
due to the difference in the momentum flow.

As was previously shown, the baryon pole contribution to the 
box-graph is the first iterate of the Lippman-Schwinger equation while these
poles do not contribute to the crossed-box graph at this order.
As in the case of the scalar exchange, the $k^0$ integration in
Eqs.~(\ref{Mboxgen}) and (\ref{MXgen}) can be performed via the contour
integration.  Whatever the explicit form of the function
$V_{\alpha\beta}(k^0, \vec{k}, q^0, \vec{q})$ its contribution to the $k^0$
integral is given by its imaginary part. The important point is that the
same function appears in ${\cal M}_{\Box}$ and ${\cal M}_{X}$. Since the
intermediate baryons can not go on shell at the meson singularities their
contribution equals to the principal values of their propagators. 

The retardation effect of the box diagram is:
\begin{eqnarray}
& {\cal M}_{\Box}^{ret}= g_{\alpha}^{2} \, g_{\beta}^{2}
\int {d^3 k\over (2\pi)^3} \int {d k^0 \over 2 \pi}
2\,Im\left[V_{\alpha\beta}(k^0, \vec{k}, q^0, \vec{q})\right] 
C^{\alpha}_{AB} C^{\beta}_{CD}
\Gamma^{A}_{\alpha (1)} (\vec{k})\Gamma^{C}_{\beta (1)} (-\vec{k}-\vec{q})
\Gamma^{B}_{\alpha (2)} (-\vec{k}) \Gamma^{D}_{\beta (2)} (\vec{k}+\vec{q}) &
\nonumber \\
& \times P\left[
{1\over (k^0 + {|\vec{p}|^2/ 2 m_B}-{|\vec{p}-\vec{k}|^2/ 2 m_B})}
{1\over (-k^0 + {|\vec{p}|^2 / 2 m_B}-{|\vec{p}-\vec{k}|^2 / 2 m_B})}\right] &
\nonumber \\
& = g_{\alpha}^{2} \, g_{\beta}^{2}
\int {d^3 k\over (2\pi)^3} \int {d k^0 \over 2 \pi} 2\, Im \left[
V_{\alpha\beta}(k^0, \vec{k}, q^0, \vec{q})  \right]  
P \left[-{1\over (k^0)^2} \right] &
\nonumber \\
& \times C^{\alpha}_{AB} C^{\beta}_{CD}
\Gamma^{A}_{\alpha (1)} (\vec{k})\Gamma^{C}_{\beta (1)} (-(\vec{k}+\vec{q}))
\Gamma^{B}_{\alpha (2)} (-\vec{k}) \Gamma^{D}_{\beta (2)} (\vec{k}+\vec{q}) \,
\left(1+{\cal O}({1\over N_c})\right) \, .&
\label{Mboxmgen}
\end{eqnarray}
Similarly, the crossed-box contribution coming entirely from the meson
singularities is:
\begin{eqnarray}
& {\cal M}_{X}= g_{\alpha}^{2} \, g_{\beta}^{2}
\int {d^3 k\over (2\pi)^3} \int {d k^0 \over 2 \pi}
2\,Im \left[V_{\alpha\beta}(k^0, \vec{k}, q^0, \vec{q})\right] 
C^{\alpha}_{AB} C^{\beta}_{CD}
\Gamma^{A}_{\alpha (1)} (\vec{k})\Gamma^{C}_{\beta (1)} (-(\vec{k}+\vec{q}))
\Gamma^{D}_{\beta (2)} (\vec{k}+\vec{q}) \Gamma^{B}_{\alpha (2)} (-\vec{k}) &
\nonumber \\
& \times P \left[ {1\over (k^0 + q^0 + {|\vec{p}|^2/ 2 m_B}-
{|\vec{p}+\vec{k}+\vec{q}|^2/ 2 m_B})
(k^0 + {|\vec{p}|^2 / 2 m_B}-{|\vec{p}-\vec{k}|^2 / 2 m_B})} \right]  &
\nonumber \\
&= g_{\alpha}^{2} \, g_{\beta}^{2}
\int {d^3 k\over (2\pi)^3} \int {d k^0 \over 2 \pi}
2\,Im \left[V_{\alpha\beta}(k^0, \vec{k}, q^0, \vec{q})\right] 
P \left[{1\over (k^0)^2} \right] & 
\nonumber \\
&\times C^{\alpha}_{AB} C^{\beta}_{CD}
\Gamma^{A}_{\alpha (1)} (\vec{k})\Gamma^{C}_{\beta (1)} (-(\vec{k}+\vec{q})) 
\Gamma^{D}_{\beta (2)} (\vec{k}+\vec{q}) \Gamma^{B}_{\alpha (2)} (-\vec{k}) \,
\left(1+{\cal O}({1\over N_c})\right)  \, &
\label{MXmgen}
\end{eqnarray}
Note that only in the nonrelativistic limit the principal values of the
baryon propagators are equal and opposite for both ${\cal M}_{\Box}^{ret}$ and
${\cal M}_{X}$. As a result, the sum of these two contributions is
proportional to a spin-flavor commutator:
\begin{eqnarray}
& {\cal M}_{\Box}^{ret}+{\cal M}_{X}
= g_{\alpha}^{2} \, g_{\beta}^{2}
\int {d^3 k\over (2\pi)^3} \int {d k^0 \over 2 \pi}
2\,Im \left[ V_{\alpha\beta}(k^0, \vec{k}, q^0, \vec{q}) \right]
P \left[-{1\over (k^0)^2} \right] &
\nonumber \\
&\times C^{\alpha}_{AB} C^{\beta}_{CD}
\Gamma^{A}_{\alpha (1)} (\vec{k})\Gamma^{C}_{\beta (1)} 
(-\vec{k}-\vec{q}) \left [\Gamma^{B}_{\alpha (2)} (-\vec{k}) \, ,\,
\Gamma^{D}_{\beta (2)} (\vec{k}+\vec{q}) \right ] 
\left(1+{\cal O}({1\over N_c})\right)\, .  &
\label{sum1}
\end{eqnarray}
In general, we have to include both orderings of the exchanged mesons. In 
Eq.~(\ref{sum1}) the first meson is $\alpha$. Changing the meson
sequence and keeping the loop momenta flow unchanged we get,
\begin{eqnarray}
& {\cal M}_{\Box}^{ret}+{\cal M}_{X}
= g_{\alpha}^{2} \, g_{\beta}^{2}
\int {d^3 k\over (2\pi)^3} \int {d k^0 \over 2 \pi}
2\, Im \left[V_{\alpha\beta}(k^0, \vec{k}, q^0, \vec{q})  \right]
P \left[-{1\over (k^0)^2} \right] &
\nonumber \\
&\times C^{\alpha}_{AB} C^{\beta}_{CD}
\left[\Gamma^{C}_{\beta (1)} (\vec{k}), 
\Gamma^{A}_{\alpha (1)} (-\vec{k}-\vec{q}) \right]
\Gamma^{D}_{\beta (2)} (-\vec{k}) 
\Gamma^{B}_{\alpha (2)} (\vec{k}+\vec{q})
\left(1+{\cal O}({1\over N_c})\right) \, ,  &
\label{sum2}
\end{eqnarray}
where we used $V_{\beta\alpha}=V_{\alpha\beta}$. Now, the commutator involves
the meson couplings along a different baryon line.

The large $N_c$ scaling of the retardation effect and the crossed-box diagram
taken separately is given by the product of the coupling constants,
$g_{\alpha}^{2} \, g_{\beta}^{2}$. However, as seen from Eqs.~(\ref{sum1}) and
(\ref{sum2}), their total contribution is proportional to the commutators
of the spin-flavor operators evaluated between the ground state baryons.
The cancelation between the retardation effect and the
crossed-box contribution up to higher order corrections happens due to the
presence of the commutator. 

A number of mesons shown in Table~\ref{tab1} and Table~\ref{tab2} make
identical
contributions to the spin-flavor structure of the nucleon-nucleon potential,
Eq.~(\ref{VNN}). For example, the $f_0$ and the time component of the 
$\omega$ contribute to the isoscalar central potential, $V_{0}^{0}$; the $\pi$
and spatial components of $\rho$ contribute to isovector tensor force. Other
such pairs are $a_0$ and $\rho^t$, $\eta$ and $\vec{\omega}$, $\pi$ and 
$\vec{\rho}$. Thus, out of ten couplings in Table~\ref{tab1} and 
Table~\ref{tab2} there are only six independent structures. They give 
thirty-six different combinations for two-meson exchange graphs counting
combinations differing in the meson sequence. Out of this the number of
distinct meson pairs is twenty-one.   

The commutators in Eqs.~(\ref{sum1}) and (\ref{sum2}) vanish identically for
those graphs in which at least one of the mesons is $f_0$ (or $\omega^t$). The
reason is that the spin-flavor structure of $f_0$ ($\omega^t$) is given by the
unity operator which commutes with any other operator. This insures the large
$N_c$ consistency of the two-meson box and crossed-box diagrams containing the
following six (independent) pairs of mesons:
\begin{equation}
(f_0 \, f_0), \,\,\, (f_0 \, \vec{f}_1), \,\,\, (f_0 \, \eta), \,\,\,
(f_0 \, a_0), \,\,\, (f_0  \, \vec{a}_1), \,\,\, (f_0 \, \pi) \, ,
\label{six}
\end{equation}
Note, as discussed above the same cancelation occurs when $\omega^t$ is
exchanged instead of $f_0$. 

Similar cancelations occur for contributions of the box and 
crossed-box diagrams containing the following pairs of mesons:
\begin{equation}
(a_0 \, \eta),\,\,\, (a_0 \, \vec{f}_{1}) \, , 
\label{two}
\end{equation}
since the spin-flavor structure of $a_0$ couplings contains only $I^a$
generators while the couplings of $\vec{f}_1$ and $\eta$ contain only $J^i$
generators which commute with $I^a$, Eq.~(\ref{SU4}).

A number of the meson-baryon couplings in Table~\ref{tab1} and Table~\ref{tab2}
are of order $N_{c}^{-1/2}$. Hence, the exchange of any pair of such mesons
is suppressed by at least $N_{c}^{-2}$ and, therefore, cannot violate the
KSM counting rules. These are the exchanges of the following meson pairs:
\begin{equation}
(a_0 \, a_0), \,\,\,(\eta \, \eta), \,\,\, (\vec{f}_1 \, \vec{f}_1), \,\,\,
(\eta \, \vec{f}_1), \,\,\, 
\left[(a_0 \, \eta), \,\,\, (a_0 \, \vec{f}_1)\right] \,,
\label{four} 
\end{equation}
where the meson pairs in the square brackets have been previously considered,
Eq.~(\ref{two}).

This leaves us with nine nontrivial meson pair exchanges whose contributions
via the box and crossed-box diagrams can potentially spoil the KSM
counting rules. Out of these, three pairs  couple to baryons only 
via non-derivative couplings:
\begin{eqnarray}
& (a_0 \, \vec{a}_1), \,\,\, (\vec{f}_1 \, \vec{a}_1),
\rightarrow {\cal O}(N_{c}^{0}) \, ,&
\nonumber \\
& (\vec{a}_1 \, \vec{a}_1),  \rightarrow {\cal O}(N_{c}^{2}) \, , &
\label{nine1}
\end{eqnarray}
and the remaining six pairs require one or two derivative couplings:
\begin{eqnarray}
& (a_0 \, \pi), \,\,\, (\eta \, \pi), \,\,\, (\eta \, \vec{a}_1) , \,\,\,
(\pi \, \vec{f}_1),  \rightarrow {\cal O}(N_{c}^{0}) \, ,&
\nonumber \\
& (\pi \, \pi), \,\,\, (\pi \, \vec{a}_1), \rightarrow {\cal O}(N_{c}^{2})\, ,&
\label{nine2}
\end{eqnarray}
In Eqs.~(\ref{nine1}) and (\ref{nine2}) the large $N_c$ scaling of the product
of corresponding coupling constants, $g_{\alpha}^{2}\,g_{\beta}^{2}$, in the
box and the crossed-box diagrams has also been indicated. 

The considerations of the meson pairs in Eq.~(\ref{nine1}) are simpler than
those with derivatively coupled mesons, Eq.~(\ref{nine2}), and will be
considered first. The analysis of the exchanges with derivative couplings
requires performing angular integration and is done in the appendix. 

The retardation effect, Eq.~(\ref{Mboxmgen}), and, the crossed-box diagram,
Eq.~(\ref{MXmgen}), involving exchanges of $a_0$ and  $\vec{a}_1$  contribute
to isoscalar and isovector spin-spin, $V_{\sigma}^{0}$ and $V_{\sigma}^{1}$,
terms; the $(\vec{f}_1\,,\vec{a}_1)$ exchange contains a $V_{0}^{1}$ term
in addition to $V_{\sigma}^{1}$. The order $N_{c}^{0}$ contributions to
$V_{\sigma}^{0}$ and $V_{0}^{0}$ from $(a_0\,,\vec{a}_1)$ and
$(\vec{f}_1\,,\vec{a}_1)$ exchanges violate the KSM rules, Eq.~(\ref{KSM}).
Fortunately, these contributions are cancelled in the sum of the retardation
effect and crossed-box diagram, Eqs.~(\ref{sum1}) and (\ref{sum2}):
\begin{eqnarray}
&C^{\vec{f}_1}_{AB} C^{\vec{a}_1}_{CD}
\Gamma^{A}_{\vec{f}_1 (1)} (\vec{k})\Gamma^{C}_{\vec{a}_1 (1)} 
(-(\vec{k}+\vec{q})) \left [\Gamma^{B}_{\vec{f}_1 (2)} (-\vec{k}) \, ,\,
\Gamma^{D}_{\vec{a}_1 (2)} (\vec{k}+\vec{q}) \right ]&
\nonumber \\
&= J^{i}_{(1)} X^{ja}_{(1)} \left[ J^{i}_{(2)} \, , \, X^{ja}_{(2)}\right]=
{1\over2} \left ( \left \{ J^{i}_{(1)} \,,\, X^{ja}_{(1)} \right\}+ 
\left[J^{i}_{(1)}\,,\, X^{ja}_{(1)}\right]\right)
\left[ J^{i}_{(2)} \,,\, X^{ja}_{(2)}\right]\,&
\nonumber \\
& = 2\, X^{ka}_{(1)}\,X^{ka}_{(2)} +{\cal O}({1\over N_{c}^{2}})\,, &
\label{f1a1}
\end{eqnarray} 
for $(\vec{f}_1\,,\vec{a}_1)$ exchange and similarly for 
$(a_0\,,\vec{a}_1)$ exchange. In the last step in Eq.~(\ref{f1a1}) we used 
the commutation and anti-commutation relations of the generators of the
contracted $SU(4)$ symmetry, Eqs.~(\ref{SU4}) and (\ref{anticom}). Thus, these
exchanges, when both box and crossed-box diagrams are included, contribute only
to the isovector spin-spin term of the nucleon-nucleon potential up to
corrections of order $N_{c}^{-2}$. This is an allowable contribution
by KSM counting rules. 

The box and crossed-box diagrams corresponding to $(\vec{a}_1,\,\vec{a}_1)$
exchange are of order $N_{c}^{2}$. The corresponding 
retardation effect and crossed-box digram separately contribute to
$V_{0}^{0}$, $V_{\sigma}^{1}$ and $V_{\sigma}^{0}$. The first two terms are of
order $N_{c}$ and the third term is of order $N_{c}^{-1}$, Eq.~(\ref{KSM}).
However, the product of the spin-flavor structures in Eqs.~(\ref{sum1}) and
(\ref{sum2}) is of order $N_{c}^{-4}$:
\begin{eqnarray}
&C^{\vec{a}_1}_{AB} C^{\vec{a}_1}_{CD}
\Gamma^{A}_{\vec{a}_1 (1)} (\vec{k})\Gamma^{C}_{\vec{a}_1 (1)} 
(-(\vec{k}+\vec{q})) \left [\Gamma^{B}_{\vec{a}_1 (2)} (-\vec{k}) \, ,\,
\Gamma^{D}_{\vec{a}_1 (2)} (\vec{k}+\vec{q}) \right ]&
\nonumber \\
&= X^{ia}_{(1)} X^{jb}_{(1)} 
\left[ X^{ia}_{(2)} \,,\, X^{jb}_{(2)}\right]= 
{1\over 2}\,\left(\left\{X^{ia}_{(1)}\,,\, X^{jb}_{(1)}\right\}+
\left [X^{ia}_{(1)}\,,\, X^{jb}_{(1)}\right ] \right)
\left[ X^{ia}_{(2)} \,,\, X^{jb}_{(2)}\right] &
\nonumber \\
&={1\over 2}\, \left [X^{ia}_{(1)}\,,\, X^{jb}_{(1)}\right ]
\left[ X^{ia}_{(2)} \,,\, X^{jb}_{(2)}\right] \sim {\cal O}({1\over N_{c}^4})
\,,&
\label{a1a1}
\end{eqnarray} 
where in the third step we used the fact that the anticommutator is symmetric
and the commutator is antisymmetric under the simultaneous exchange of the
spin-flavor indices $(ia) \rightarrow (jb)$; the large $N_c$ of the baryon
matrix elements of $\left[ X^{ia} \,,\, X^{jb}\right]$ is given in 
Eq.~(\ref{X}). Combining the $N_{c}^{4}$ suppression in Eq.~(\ref{a1a1}) with
the $N_{c}^{2}$ scaling of the product $g_{\vec{a}_1}^{4}$ we see that the
sum in Eq.~(\ref{sum1}) (and similarly in Eq.~(\ref{sum2})) is of order
$N_{c}^{-2}$ which is consistent with KSM counting rules. Note that in this
case full contracted $SU(4)$ algebra has to be used to insure the cancelation.
Thus, the cancelation of the retardation effect against the crossed-box diagram
requires an inclusion of both nucleon and $\Delta$ intermediate states. If one
restricts the intermediate states to nucleons only, the cancelation would not
occur.

Thus far, we have shown that the retardation effect of all two-meson
exchange diagrams without derivative couplings, Eq.~(\ref{nine1}), cancel 
against the corresponding crossed-box graphs. As is shown in the appendix, 
similar cancelations occur for the remaining six meson pairs, 
Eq.~(\ref{nine2}), which involve one or two derivatively coupled mesons.
 
In addition to box and crossed-box diagrams, any pair of mesons can be
exchanged via triangle (or ``seagull) diagrams, Fig.~\ref{fig6},
containing a 4-point meson-baryon vertex. The spin-flavor structure of this
vertex is given by the product of  two $\Gamma^{A}_{\alpha (n)}(\vec{k})$
operators. The 4-point meson-baryon coupling is of order $N_{c}^{0}$
for any meson pair \cite{LN2}. Hence, the largest scaling of a triangle
diagram is $N_c$, {\it e.g.} when two pions or $f_0$ and $\omega^t$
are exchanged. Thus, the triangle graph can not violate the scaling of the
leading $({\cal O}(N_c))$ spin-flavor terms, Eq.~(\ref{KSM}). However, the 
subleading $({\cal O}(N_{c}^{-1}))$ terms might be sensitive to contributions
from the triangle diagrams. Since these diagrams contain only one baryon 
propagator its pole does not contribute to the potential (the contour of the
complex $k^0$ integration can always be closed in such a way as to avoid the
baryon pole). As we will show shortly, the contributions from the meson 
singularities in the triangle graphs add up to cancel all terms that violate
the KSM counting rules.

A given triangle graph can be associated
with a corresponding box or crossed-box diagram by shrinking the appropriate
baryon propagator to zero. It can then be shown that the sum of the
appropriate pair of the triangle graphs, shown in Fig.~\ref{fig6}, is similar
to Eq.~(\ref{sum1}) or Eq.~(\ref{sum2}). The essential point in the above
discussion was the presence of the commutator in Eqs.~(\ref{sum1}) and
(\ref{sum2}). The same commutator appears in the sum of the triangle graphs. 

What are the pairs of the triangle graphs that correspond to the box and 
crossed-box diagrams which led to Eqs.~(\ref{sum1}) and (\ref{sum2})? These 
graphs contain the same meson pairs. The two corresponding graphs differ
according to which baryon line the 4-point meson-baryon vertex is attached,
Fig.~\ref{fig6}. In addition, two $\Gamma$ structures at the 4-point vertex of
one of the corresponding graphs are in opposite order relative to the sequence
of these structures in the other graph. This leads to the appearance of a
commutator in the sum of the corresponding triangle graphs. 

In the case of the box and crossed-box diagrams the direction of the energy
flow assured that the retardation effect and the crossed-box contribution are
equal and opposite up to $1/N_c$ corrections. The sign
difference was due to the product of the principle values of the baryon
propagators (after the nonrelativistic reduction), Eqs.~(\ref{Mboxmgen}) and
(\ref{MXmgen}), which had different signs for the box and crossed-box
diagrams. Despite the presence of only a single baryon propagator, the
contributions from each of the corresponding triangle graphs come with
opposite signs due to the different flow of the energy and momenta,
Fig.~\ref{fig6}. The sum of these two graphs is:
\begin{eqnarray}
& {\cal M}_{1}+{\cal M}_{2}
= g_{\alpha} \, g_{\beta}\, g_{\alpha\beta}
\int {d^3 k\over (2\pi)^3} \int {d k^0 \over 2 \pi}
2\, Im \left[V_{\alpha\beta}(k^0, \vec{k}, q^0, \vec{q}) \right]
P \left[-{1\over k^0} \right] &
\nonumber \\
&\times C^{\alpha}_{AB} C^{\beta}_{CD}
\Gamma^{A}_{\alpha (1)} (\vec{k})\Gamma^{C}_{\beta (1)} 
(-\vec{k}-\vec{q}) \left [\Gamma^{B}_{\alpha (2)} (-\vec{k}) \, ,\,
\Gamma^{D}_{\beta (2)} (\vec{k}+\vec{q}) \right ] 
\left(1+{\cal O}({1\over N_c})\right)\, ,  &
\label{triangle}
\end{eqnarray}   
where $g_{\alpha\beta}$ (order $N_{c}^{0}$ for all $\alpha$ and $\beta$) is 
the coupling constant of the 4-point vertex and the function $V_{\alpha\beta}$
is given by Eq.~(\ref{Valpha}) provided $\tilde{V}_{\alpha\beta}$ contains
the product of the three meson baryon vertex functions (including one
corresponding to 4-point vertex). A similar expression can be written for the
sum of the two triangle graphs in which the sequence of the $\alpha$ and
$\beta$ mesons is changed as in Eq.~(\ref{sum2}).

It is clear that the sum in Eq.~(\ref{triangle}) contributes to the same
spin-flavor terms as the sum in Eq.~(\ref{sum1}): both expressions contain
identical spin-flavor structures. The differences in the integrands are 
irrelevant as far as the cancelations in Eqs.~(\ref{sum1}) and (\ref{sum2})
are concerned. As a result, the large $N_c$ scaling of the contribution in
Eq.~(\ref{triangle}) is that of Eq.~(\ref{sum1}) times scaling of 
$(g_{\alpha}\,g_{\beta})^{-1}$.

Hence, when all the contributions of the triangle graphs are included the
resulting spin-flavor terms are consistent with the KSM counting rules, 
Eq.~(\ref{KSM}).

\section{Conclusion}

At a technical level we have shown by explicit calculation that if
meson-baryon couplings scale according to the standard large $N_c$
rules, then the two-meson-exchange contributions to the nonrelativistic
baryon-baryon potential is consistent with the large $N_c$ KSM scaling rules
deduced in Refs.~\cite{NN1,NN2}. This is highly nontrivial since the
derivation of these rules in Refs.~\cite{NN1,NN2} only included diagrams which
correspond to one-meson-exchange when translated to the hadronic level. This
certainly adds confidence that Refs.~\cite {NN1,NN2} correctly described the
$N_c$ scaling behavior of the nucleon-nucleon potential in the large $N_c$
limit of QCD. The essential issue in the calculations here was that the
retardation contributions to the potential from the box graph cancel against
the crossed-box contributions for all spin-isospin structures in the potential
where the retardation contributions or the crossed-box contributions separately
violate the counting rules. 

The derivation presented here was done in a ``brute force'' manner. Namely,
we considered the various meson exchanges one at a time, identified the
contributions to the various spin-isospin structures which apparently
violated the KSM large $N_c$ scalings and showed that in all cases they
canceled. It would be very useful to find a more general method for
demonstrating the cancelation. While the methods used here were adequate
for the two-meson exchange case, it would be extremely cumbersome to
extend them to three-meson exchange or higher. Given the cancelations for
all ``dangerous'' contributions at the two-meson-exchange level it seems
reasonable to expect that such cancelations will occur for any number of
meson exchanges and that the full baryon-baryon potential will be
consistent with the KSM scaling rules. However, a general proof of the
cancelations for all orders would be desirable.

In the Introduction it was argued that the large $N_c$ scaling behavior of
the baryon-baryon interactions give some general insights into the
underlying physics arising from QCD. In particular it was argued that a
consistent picture emerged and five aspects of this picture were
enumerated. Let us now briefly discuss how the calculations discussed
above support this picture.

The first point raised was that while a nonrelativistic potential used to
describe the interaction has overall strength of order $N_c$, the kernel
of a Bethe-Salpeter equation does not have a simple $N_c$ dependence. As
noted many times, the order $N_{c}^{2}$ contribution to the potential from
the crossed-box graph is canceled by the retardation effect from the box
graph. However, such a cancelation cannot happen in the context of the
Bethe-Salpeter equation. The entire box graph (including meson pole
contributions) is an iterate of the Bethe-Salpeter kernel and hence cannot
be included as a contribution to the kernel. Thus, in the Bethe-Salpeter
context there is no part of the box graph to cancel the order $N_{c}^2$
contribution from the crossed-box.  Therefore, unlike the potential, the
Bethe-Salpeter kernel cannot be associated with an overall strength of
$N_c$. Presumably, the Bethe-Salpeter kernel has contributions scaling as
$N_c$ to all powers arising from multiple meson exchanges.

The second point made was that the meson exchange
picture of baryon-baryon interactions with the leading part of the
potential scaling as $N_c$ is consistent with the meson exchange picture
of the potential provided the momentum exchanged is of order $N_{c}^{0}$.
It was shown explicitly using nonrelativistic kinematics that at the
level of two-meson exchange all ``dangerous'' contributions to the
potential canceled so that there is no inconsistency with a potential
scaling as $N_c$. It is reasonable to expect the behavior to hold for any
number of meson exchanges. If true, this strengthens the case for using
meson exchange models to describe nucleon-nucleon interactions.

However, it was also argued that the idea of a potential described by the
meson exchange picture is unsuitable for momenta of order $N_c$. At a
technical level this is apparent in Eqs.~(\ref{MBm}), (\ref{MX}),
(\ref{MXgen}) and (\ref{Mboxmgen}) where the cancelations of the box and
crossed box graphs depend explicitly on the nonrelativistic form of the
propagator. If momenta of order $N_c$ were used the cancelations clearly
fail to occur. Thus the usefulness of the meson-exchange picture is not
evident for momenta of order $N_c$. In fact, it is quite satisfying that
the evidence of consistency breaks down in this regime for a number of
reasons.  In the first place Witten's TDH picture of baryon-baryon
interactions is more appropriate for $p \sim N_c$. This picture has no
obvious meson-exchange interpretation. The internal structure of each
baryon is distorted in the presence of the other.  Moreover, it is not
surprising from a more traditional hadronic viewpoint that a
meson-exchange potential picture breaks down in this regime. If $p \sim
N_c$ and $m_N \sim N_c$ then the kinetic energy of the baryons is also of
order $N_c$.  Since meson masses are of order $N_c^0$ an increasing number of
mesons are produced. It is hardly surprising that the potential picture breaks
down in this situation.
				
A fourth point raised in the Introduction was that relative sizes of the
various spin-isospin structures in the nucleon-nucleon potential are
consistent at the two-meson-exchange level with those deduced from the
contracted $SU(4)$ structure of KSM. Moreover, if one looks carefully at
all of the cancelations, one finds that corrections to the leading
behavior were all $1/N_{c}^{2}$ suppressed.  This is consistent with
Refs.~\cite{NN1,NN2} where it is found that subleading spin-isospin
structures are down by factors $1/N_{c}^{2}$.  Overall this strongly
supports the view that the expansion is in fact in $1/N_{c}^{2}$ rather
than in $1/N_c$.

The final point stressed was that the $\Delta$ plays
an essential role.  As is evident from Eqs.~(\ref{a1a1}) and the appendix,
the cancelations between the box and crossed-box graphs do not occur if
intermediate states are restricted entirely to nucleons; $\Delta$
resonances are required as intermediate states. More generally one expects
that as the contracted $SU(4)$ structure is used to obtain cancelations
the entire $I=J=1/2, 3/2, 5/2, ...$ tower of baryon states can contribute.
Up to two-meson exchange with nucleon as initial and final states,
however, only nucleons and $\Delta$'s can contribute.

The formal consistency of the large $N_c$ treatment and the meson exchange
picture is quite satisfying.  However, considerable caution should be
exercised in trying to draw conclusions about the real world of $N_c=3$.
We have used $1/N_c$ as a counting parameter to distinguish large from
small contributions.  This is clearly legitimate if all the coefficients
multiplying these factors are natural, {\it i.e.} of order unity.  
However, all coefficients are not natural.  One key difficulty is that the
meson-exchange picture is being used here to connect hadronic phenomena
with nuclear phenomena.  However, the scales in nuclear physics are
generally much smaller than those in hadronic physics \cite{SF3}. It is
not clear directly from QCD why these nuclear scales are so small and it
is generally thought to be ``accidental''. The interplay between small
nuclear scales (that may be large in a $1/N_c$ sense) with much larger
hadronic scales (that may be small formally in a $1/N_c$ sense) can
potentially spoil the results of a straightforward $1/N_c$ approach. To
show how extreme this problem may be we can consider the deuteron binding
energy, {\it B} (which is formally of order $N_c$) and the delta-nucleon
mass difference, $m_{\Delta}-m_{N}$ (which is order $1/N_c$). If all
coefficients were natural, one would expect $B$ to be an order of
magnitude larger than $m_{\Delta}-m_{N}$, whereas, in fact, it is two orders
of magnitude smaller. It would not be surprising that difficulties might
arise when calculating $B$ if one neglects $m_{\Delta}-m_{N}$ as being
``small''.

The large $N_c$ structure of the nucleon-nucleon potential has been so far
used phenomenologically in two contexts. The first is as an attempt to
justify the observed approximate Wigner $SU(4)$ symmetry \cite{Wigner} 
in light nuclei
as arising from the underlying contracted $SU(4)$ structure in the large
$N_c$ potential \cite{NN1}. The second is an attempt to justify the
qualitative sizes of the spin-flavor structures in phenomenological
potentials as being explained by contracted $SU(4)$ structure in the large
$N_c$ potential \cite{NN1,NN2}. It is not immediately obvious that these
two explanations are legitimate in light of qualitatively distinct nuclear
scales that are not associated with the $1/N_c$ expansion. Clearly, this
issue needs further investigation. However, it is also not immediately
clear how to formulate a systematic expansion which both incorporates the
$1/N_c$ scaling rules while allowing nuclear scales to be much smaller
than hadronic scales. The comparison of the qualitative sizes of the
spin-flavor structures in phenomenological potentials with what is
expected from large $N_c$ raises another issue.  The potentials predicted
in large $N_c$ are not nucleon-nucleon potentials; rather, they are coupled
channel potentials for the full tower of $I=J$ baryon states including an
explicit $\Delta$. The phenomenological potentials to which they are
compared have the explicit $\Delta$'s integrated out.  It is by no means
clear that the act of integrating out $\Delta$'s does not alter the
spin-flavor structure.  Again, this requires further study.

Given these possible difficulties in drawing phenomenological conclusions
from large $N_c$ potentials, one might ask about the relevance for the
real world of our demonstration that large $N_c$ counting rules are
consistent with the meson-exchange picture of potentials at the
two-meson-exchange level. Of course, it remains possible that after a
careful study one may find that the particular phenomenological
predictions to date---the Wigner $SU(4)$ symmetry and the characteristic
relative sizes of the various terms in phenomenological potentials---are
robust and remain valid even after the smallness of the typical nuclear
scales are included. Whether or not this turns out to be the case,
however, we may still be able to learn qualitatively interesting things.
For example, the cancelations seen in the two-pion exchange graphs require
that $\Delta$ intermediate states be included. For $N_c=3$, one does not
expect such cancelations to be perfect, but the general tendency to cancel
should survive. This suggests that $\Delta$ box and crossed-box
contributions should be comparable in size to the ones with nucleon
intermediate states.  This issue may be relevant for potential models
motivated by chiral symmetry where two-pion exchange contributions with
nuclear intermediate states are included at next-to-leading order but
explicit $\Delta$ contributions are not included\cite{SF3}. At a more
qualitative level, the fact that at large $N_c$ a meson-exchange motivated
picture of the potential is consistent gives at least some support for the
view that more generally nucleon-nucleon interactions can be described in
terms of meson exchanges.

\acknowledgements

This work is supported by the U.S.~Department of Energy grant 
DE-FG02-93ER-40762. TDC wishes to acknowledge discussions with M.~Luty,
D.~Phillips, J.~Friar, F.~Gross and S.~Wallace. He also acknowledges support
at the INT where progress was made on this work.

\appendix
\section*{}

In this appendix we discuss the box and crossed-box graphs with one or two
derivative couplings, Eq.~(\ref{nine2}). As in the case of the
non-derivatively coupled mesons, Eq.~(\ref{nine1}),
the sum of the retardation effect and the crossed-box diagram can be written as
the sum of the products of anticommutators and commutators of the spin-flavor
generators. The cancelation of the terms that violated spin-flavor counting
rules essentially has occurred due the symmetry properties of this products
under the interchange of the spin-flavor symmetry. The remaining terms are
either consistent with the counting rules or suppressed by $1/N_{c}^{2}$ as in
Eq.~(\ref{a1a1}). However, when derivatively coupled mesons are included
the symmetry of the products of commutators and anticommutators under
the interchange of spin-flavor symmetry is broken due to the contraction of 
the spin-flavor generators with momentum indices. In order to see the
cancelation the angular integration in Eqs.~(\ref{sum1}) and (\ref{sum2})
must be performed. In this appendix the cancelation is shown for the
two-pion exchange diagrams. The exchanges involving other pairs of mesons,
Eq.~(\ref{nine2}), are essentially identical to this case.

The retardation effect and the crossed-box contribution is given in 
Eq.~(\ref{sum1}) which has the following form for the two-pion exchange:
\begin{eqnarray}
& {\cal M}_{\Box}^{ret}+{\cal M}_{X}
= g_{\pi}^{4} \int {d^3 k\over (2\pi)^3} \int {d k^0 \over 2 \pi}
P \left[-{1\over (k^0)^2} \right] 
2\, Im \left[V_{\pi\pi}(k^0, \vec{k}, q^0, \vec{q}) \right] &
\nonumber \\
&\times 
X^{ia}_{(1)}\,X^{jb}_{(1)} \left[ X^{la}_{(2)} \,,\, X^{rb}_{(2)}\right] \,
\vec{k}^{i} \vec{k}^{l} (\vec{k}^{j}+\vec{q}^{j}) (\vec{k}^{r}+\vec{q}^{r})
\left(1+{\cal O}({1\over N_c})\right) \,.&
\label{pipi1}
\end{eqnarray}
Performing $k^0$ integration Eq.~(\ref{pipi1}) reduces to
\begin{equation}
{\cal M}_{\Box}^{ret}+{\cal M}_{X}
= g_{\pi}^{4} \int {k^2 \, d k\over (2\pi)^3} \int d \Omega
\,\, F(|\vec{k}|, |\vec{q}|, \vec{k}\cdot\vec{q}, q^0) \,
X^{ia}_{(1)}\,X^{jb}_{(1)} \left[ X^{la}_{(2)} \,,\, X^{rb}_{(2)}\right] \,
\vec{k}^{i} \vec{k}^{l} (\vec{k}^{j}+\vec{q}^{j}) (\vec{k}^{r}+\vec{q}^{r})\,,
\label{pipi2}
\end{equation}
where the explicit form of the function 
$F(|\vec{k}|, |\vec{q}|, \vec{k}\cdot\vec{q}, q^0)$ is not required for the
following discussion.

Let us consider the angular integration in Eq.~(\ref{pipi2}):
\begin{equation}
I_{\Omega} \equiv 
\int d \Omega \,\, F(|\vec{k}|, |\vec{q}|, \vec{k}\cdot\vec{q}, q^0) \,\,\,
\vec{k}^{i} \vec{k}^{l} (\vec{k}^{j}+\vec{q}^{j}) (\vec{k}^{r}+\vec{q}^{r})\,.
\label{pipi3}
\end{equation}
The general form of this integral is:
\begin{eqnarray}
& I_{\Omega} = \left( 
\delta^{ij}\,\delta^{lr}+\delta^{il}\,\delta^{jr}+\delta^{ir}\,\delta^{jl} 
\right) \, f_{1}(|\vec{k}|, |\vec{q}|, q^0)+ 
q^i\, q^j\, q^l\, q^r \, f_{2}(|\vec{k}|, |\vec{q}|, q^0) &
\nonumber \\
& + \left ( \delta^{il}\, q^j\, q^r + \delta^{jr}\, q^i\, q^l \right)
f_{3}(|\vec{k}|, |\vec{q}|, q^0) 
+ \left (\delta^{ij}\, q^l\, q^r + \delta^{ir}\, q^j\, q^l + 
\delta^{jl}\, q^i\, q^r + \delta^{lr}\, q^i\, q^j \right )
f_{4}(|\vec{k}|, |\vec{q}|, q^0) \,,&
\label{pipi4}
\end{eqnarray} 
where functions $f_1$, $f_2$, $f_3$ and $f_4$ do not depend on
$\vec{k}\cdot\vec{q}$. The form of the $I_{\Omega}$ can be obtained from 
general arguments based on the symmetry properties of Eq.~(\ref{pipi3})
under the various exchanges of the momentum indices.

Each term in Eq.~(\ref{pipi4}) when combined with spin-flavor generators in
Eq.~(\ref{pipi1}) is suppressed by $N_{c}^{-4}$. Let us see how it comes about
for each term separately. 

The first product of Kronecker deltas in the $f_1$ term leads to:
\begin{eqnarray}
& \delta^{ij}\,\delta^{lr} X^{ia}_{(1)}\,X^{jb}_{(1)} 
\left[ X^{la}_{(2)} \,,\, X^{rb}_{(2)}\right]=
X^{ia}_{(1)}\,X^{ib}_{(1)} \left[ X^{la}_{(2)} \,,\, X^{lb}_{(2)}\right] 
={1\over 2} \left(\left\{X^{ia}_{(1)}\,,\, X^{ib}_{(1)}\right\}+
\left [X^{ia}_{(1)}\,,\, X^{ib}_{(1)}\right ] \right)
\left[ X^{la}_{(2)} \,,\, X^{lb}_{(2)}\right] &
\nonumber \\
&={1\over 2} \left\{X^{ia}_{(1)}\,,\, X^{ib}_{(1)}\right\}
\left[ X^{la}_{(2)} \,,\, X^{lb}_{(2)}\right] 
+{1\over 2}
\left [X^{ia}_{(1)}\,,\, X^{ib}_{(1)}\right ] 
\left[ X^{la}_{(2)} \,,\, X^{lb}_{(2)}\right] 
= {\cal O}({1\over N_{c}^{4}})\,, &
\label{f11}
\end{eqnarray}
where the product of the commutator and the anticommutator vanishes because
the commutator is antisymmetric under the  $a \leftrightarrow b$ exchange
while the anticommutator is symmetric. The remaining product of the two
commutators is of order $N_{c}^{-4}$ from Eq.~(\ref{XX}). Thus, an overall 
contribution of this term is of order $N_{c}^{-2}$ which is consistent with
KSM counting rules, Eq.~(\ref{KSM}). Note how the angular averaging of 
Eq.~(\ref{pipi3}) induced symmetry properties of the spin-flavor
matrices in Eq.~(\ref{pipi1}).

The second product of deltas in $f_1$ leads to the expression identical to
Eq.~(\ref{a1a1}); its contribution, therefore, is suppressed by $N_{c}^{-4}$
as well. The suppression of the last term multiplying the $f_1$ are easily 
observed:
\begin{eqnarray}
& \delta^{ir}\,\delta^{jl} X^{ia}_{(1)}\,X^{jb}_{(1)} 
\left[ X^{la}_{(2)} \,,\, X^{rb}_{(2)}\right]=
X^{ia}_{(1)}\,X^{ib}_{(1)} \left[ X^{ja}_{(2)} \,,\, X^{ib}_{(2)}\right] 
={1\over 2} \left(\left\{X^{ia}_{(1)}\,,\, X^{jb}_{(1)}\right\}+
\left [X^{ia}_{(1)}\,,\, X^{jb}_{(1)}\right ] \right)
\left[ X^{ja}_{(2)} \,,\, X^{ib}_{(2)}\right] &
\nonumber \\
&={1\over 2} \left\{X^{ia}_{(1)}\,,\, X^{jb}_{(1)}\right\}
\left[ X^{ja}_{(2)} \,,\, X^{ib}_{(2)}\right] 
+{\cal O}({1\over N_{c}^{4}})= {\cal O}({1\over N_{c}^{4}}) \,,&
\label{f13a}
\end{eqnarray}
where in the last step we used antisymmetry under the simultaneous exchange of
$a \leftrightarrow b$ and $i \leftrightarrow j$:
\begin{equation}
\left\{X^{ia}_{(1)}\,,\, X^{jb}_{(1)}\right\}
\left[ X^{ja}_{(2)} \,,\, X^{ib}_{(2)}\right]=
- \, \left\{X^{ia}_{(1)}\,,\, X^{jb}_{(1)}\right\}
\left[ X^{ja}_{(2)} \,,\, X^{ib}_{(2)}\right] \,.
\label{f13b}
\end{equation}

Similarly the product of the 4 components of the external momenta multiplying 
$f_2$ is of order $N_{c}^{-4}$:
\begin{equation}
 q^i \,q^j \,q^l \,q^r \, X^{ia}_{(1)}\,X^{jb}_{(1)} 
\left[ X^{la}_{(2)} \,,\, X^{rb}_{(2)}\right] 
 ={1\over 2} \, q^i \,q^j \,q^l \,q^r \,
\left\{X^{ia}_{(1)}\,,\, X^{jb}_{(1)}\right\}
\left[ X^{la}_{(2)} \,,\, X^{rb}_{(2)}\right] + {\cal O}({1\over N_{c}^{4}})=
{\cal O}({1\over N_{c}^{4}}) \,, 
\label{f21a}
\end{equation}
where the vanishing of the last product can be seen after the substitution
$a \leftrightarrow b$, $i \leftrightarrow j$ and $l \leftrightarrow r$:
\begin{equation}
q^i \,q^j \,q^l \,q^r \, \left\{X^{ia}_{(1)}\,,\, X^{jb}_{(1)}\right\}
\left[ X^{la}_{(2)} \,,\, X^{rb}_{(2)}\right] =
-q^i \,q^j \,q^l \,q^r \, \left\{X^{ia}_{(1)}\,,\, X^{jb}_{(1)}\right\}
\left[ X^{la}_{(2)} \,,\, X^{rb}_{(2)}\right] \, .
\label{f21b}
\end{equation}

Similar arguments can be used to show the vanishing
(up to ${\cal O}(N_{c}^{-4})$) of the term containing the function $f_{3}$:
\begin{eqnarray} 
& \left (\delta^{il}\, q^j\, q^r + \delta^{jr}\, q^i\, q^l \right)
\, X^{ia}_{(1)}\,X^{jb}_{(1)} \left[ X^{la}_{(2)} \,,\, X^{rb}_{(2)}\right] &
\nonumber \\
& = q^j \, q^r \,
X^{ia}_{(1)}\,X^{jb}_{(1)} \left[ X^{ia}_{(2)} \,,\, X^{rb}_{(2)}\right]+
q^i \, q^l \,
X^{ia}_{(1)}\,X^{jb}_{(1)} \left[ X^{la}_{(2)} \,,\, X^{jb}_{(2)}\right] &
\nonumber \\
& = q^j \, q^l \,
X^{ia}_{(1)}\,X^{jb}_{(1)} \left[ X^{ia}_{(2)} \,,\, X^{lb}_{(2)}\right]+
q^j \, q^l \,
X^{ja}_{(1)}\,X^{ib}_{(1)} \left[ X^{la}_{(2)} \,,\, X^{ib}_{(2)}\right] &
\nonumber \\
& = q^j \, q^l \,
X^{ia}_{(1)}\,X^{jb}_{(1)} \left[ X^{ia}_{(2)} \,,\, X^{lb}_{(2)}\right] +
q^j \, q^l \,
X^{jb}_{(1)}\,X^{ia}_{(1)} \left[ X^{ia}_{(2)} \,,\, X^{lb}_{(2)}\right] &
\nonumber \\
& = q^j \, q^l \,\left[ X^{ia}_{(1)} \,,\, X^{jb}_{(1)}\right] \,
\left[ X^{ia}_{(2)} \,,\, X^{lb}_{(2)}\right] = {\cal O}({1\over N_{c}^{4}})
\, , &
\label{f3}
\end{eqnarray}
where in the second equality we changed the index $r$ into $l$ in the first 
term and exchanged $i \leftrightarrow j$ in the second term; in the next step
we make the $a \leftrightarrow b$ exchange in the second term. 

The first term multiplying $f_4$ in Eq.~(\ref{pipi4}) is suppressed as follows:
\begin{eqnarray}
& \delta^{ij}\,q^l \, q^r \, X^{ia}_{(1)}\,X^{jb}_{(1)} 
\left[ X^{la}_{(2)} \,,\, X^{rb}_{(2)}\right]=
q^l \, q^r \,
X^{ia}_{(1)}\,X^{ib}_{(1)} \left[ X^{la}_{(2)} \,,\, X^{rb}_{(2)}\right] &
\nonumber \\
& ={1\over 2}\,q^l \, q^r \,
\left \{ X^{ia}_{(1)}\,,\, X^{ib}_{(1)} \right\} 
\left[ X^{la}_{(2)} \,,\, X^{rb}_{(2)}\right] +
{\cal O}({1\over N_{c}^{4}}) ={\cal O}({1\over N_{c}^{4}})  &
\label{f4a}
\end{eqnarray}
where the vanishing of the last product can be seen via the substitution 
$a \leftrightarrow b$ followed by $l \leftrightarrow r$. Similar arguments
apply for the last term multiplying $f_{4}$.

Lastly, the sum of the remaining two terms in Eq.~(\ref{pipi4}) multiplying
$f_4$ vanishes (up to ${\cal O}(N_{c}^{-4})$) as follows:
\begin{eqnarray}
& \left (\delta^{ir}\, q^j\, q^l + \delta^{jl}\, q^i\, q^r \right )
\, X^{ia}_{(1)}\,X^{jb}_{(1)} \left[ X^{la}_{(2)} \,,\, X^{rb}_{(2)}\right] 
 = q^j \, q^l \,
X^{ia}_{(1)}\,X^{jb}_{(1)} \left[ X^{la}_{(2)} \,,\, X^{ib}_{(2)}\right] +
q^i \, q^r \,
X^{ia}_{(1)}\,X^{jb}_{(1)} \left[ X^{ja}_{(2)} \,,\, X^{rb}_{(2)}\right] &
\nonumber \\
& = q^i \, q^l \,
X^{ja}_{(1)}\,X^{ib}_{(1)} \left[ X^{la}_{(2)} \,,\, X^{jb}_{(2)}\right]+
q^i \, q^l \,
X^{ia}_{(1)}\,X^{jb}_{(1)} \left[ X^{ja}_{(2)} \,,\, X^{lb}_{(2)}\right] &
\nonumber \\
& = q^i \, q^l \,
X^{ja}_{(1)}\,X^{ib}_{(1)} \left[ X^{la}_{(2)} \,,\, X^{jb}_{(2)}\right]+
q^i \, q^l \,
X^{ib}_{(1)}\,X^{ja}_{(1)} \left[ X^{jb}_{(2)} \,,\, X^{la}_{(2)}\right] &
\nonumber \\
&= {1\over 2}\, q^i \, q^l \,
\left \{ X^{ja}_{(1)}\,,\, X^{ib}_{(1)} \right \}
\left[ X^{la}_{(2)} \,,\, X^{jb}_{(2)}\right]+
{1\over 2}\,q^i \, q^l \,
\left \{ X^{ib}_{(1)}\,,\, X^{ja}_{(1)} \right \}
\left[ X^{jb}_{(2)} \,,\, X^{la}_{(2)}\right] + {\cal O}({1\over N_{c}^{4}}) =
{\cal O}({1\over N_{c}^{4}}) \, , &
\label{f4b}
\end{eqnarray}
where in the second equality we changed $i \leftrightarrow j$ $i$ and 
$r \leftrightarrow l$ in the first and second term respectively; 
in the next step the change is $a \leftrightarrow b$ in the second term.

This completes the discussion of two-pion exchange box and crossed-box 
diagrams. Their mutual contribution has an overall scaling of $N_{c}^{-2}$ and
is, therefore, consistent with the KSM counting rules, Eq.~(\ref{KSM}).


\begin{references}

\bibitem{LN1} G.~'t~Hooft, Nucl.~Phys.~{\bf B72} 461 (1974).  
\bibitem{LN2} E.~Witten, Nucl.~Phys.~{\bf B160} 57 (1979).
\bibitem{SF1} J.L.~Gervais and B. Sakita, Phys.~Rev.~Lett.~{\bf 52} 87 (1984); 
\\
J.L.~Gervais and B. Sakita, Phys.~Rev.~{\bf D30} 1795 (1984); \\
C.~Carone, H.~Georgi, S.~Osofsky, Phys.~Lett.~{\bf B322} 227 (1994); \\ 
M.~Luty and J.~March-Russell, Nucl.~Phys.~{\bf B426} 71 (1994).
\bibitem{SF2} R.~Dashen, E.~Jenkins and A.V.~Manohar, Phys.~Rev.~{\bf D49} 
4713 (1994).  
\bibitem{SF3} R.~Dashen, E.~Jenkins and A.V.~Manohar, Phys.~Rev.~{\bf D51} 
3697 (1995).
\bibitem{NN1} D.B.~Kaplan and M.J.~Savage, Phys.~Lett.~{\bf B365} 244 (1996).
\bibitem{NN2} D.B.~Kaplan and A.V.~Manohar, Phys.~Rev.~{\bf C56} 76 (1997).  
\bibitem{MNN} R.~Machleidt, K.~Holinde and Ch.~Elster, Phys.~Rep.~{\bf 149} 1
(1987).
\bibitem{PhNN} M.M.~Nagels, T.A.~Rijken and J.J.~de Swart,
Phys.~Rev.~{\bf D17} 768 (1978); \\
R.~Wiringa, R.~Smith and T.~Ainsworth, Phys.~Rev.~{\bf C29} 1207 (1984).
\bibitem{SK} G.S.~Adkins, C.R.~Nappi and E.~Witten, Nucl.~Phys.~{\bf B228} 552
(1983).
\bibitem{cancel} F.~Gross, Phys.~Rev.~186 1448 (1969);
F.~Gross, Phys.~Rev.~{\bf C26} 2203 (1982); \\
F.~Gross, Relativistic Quantum Mechanics and Field Theory (Wiley, 1993).
\bibitem{Wigner} E.~Wigner, Phys.~Rev.~{\bf 51} 106, 947 (1937); 
{\it ibid.} {\bf 56} 519 (1939).
\bibitem{chNN} S.~Weinberg, Phys.~Lett.~{\bf B251} 288 (1990); 
S.~Weinberg, Nucl.~Phys.~{\bf B363} 3 (1991); \\
C.~Ord\'o\~nez and U. van~Kolck, Phys.~Lett.~{\bf B291} 459 (1992); \\
C.~Ord\'o\~nez, L.~Ray and U. van~Kolck, Phys.~Rev.~Lett.~{\bf 72} 1982
(1994);\\
E.~Epelbaoum, W.~Gl\"ockle and Ulf-G.~Mei\ss ner,  Nucl.~Phys.~{\bf A637} 107
(1998); \\ 
E.~Epelbaoum, W.~Gl\"ockle and Ulf-G.~Mei\ss ner, Nucl.~Phys.~{\bf A671} 295
(2000).


\end{references}
\end{document}